\documentclass[a4paper,12pt]{article}
\pdfoutput=1

\usepackage{jheppub,fancyhdr,graphicx,epstopdf,color,amsmath,cases,slashed,hyperref,multirow}
\usepackage{cancel}
\usepackage{float}

\newcommand{\beq}{\begin{eqnarray}}% can be used as {equation} or {eqnarray}
\newcommand{\eeq}{\end{eqnarray}}

\def\ltap{\ \raise.3ex\hbox{$<$\kern-.75em\lower1ex\hbox{$\sim$}}\ }
\def\gtap{\ \raise.3ex\hbox{$>$\kern-.75em\lower1ex\hbox{$\sim$}}\ }

\def\be{\begin{equation}}
\def\ee{\end{equation}}
\def\bea{\begin{eqnarray}}
\def\eea{\end{eqnarray}}

\definecolor{red1}{cmyk}{0,1,1,0.3}

\title{\boldmath $hh+\text{jet}$ production at 100 TeV}

\author[a,b]{Shankha~Banerjee}
\author[c]{\!\!, Christoph~Englert}
\author[d]{\!\!, Michelangelo~L.~Mangano}
\author[d]{\!\!, Michele~Selvaggi}
\author[a]{\!\!, Michael~Spannowsky}

\affiliation[a]{Institute for Particle Physics Phenomenology, Department of Physics, Durham University, Durham DH1 3LE, United Kingdom}
\affiliation[b]{Universit\'{e} Grenoble Alpes, USMB, CNRS, LAPTh, F-74000 Annecy, France}
\affiliation[c]{School of Physics and Astronomy, University of Glasgow, Glasgow, G12 8QQ, United Kingdom}
\affiliation[d]{CERN, CH-1211 Geneva, Switzerland}

\emailAdd{shankha.banerjee@durham.ac.uk}
\emailAdd{christoph.englert@glasgow.ac.uk}
\emailAdd{michelangelo.mangano@cern.ch}
\emailAdd{michele.selvaggi@cern.ch}
\emailAdd{michael.spannowsky@durham.ac.uk}

\abstract
{Higgs pair production is a crucial phenomenological process in deciphering the nature of the TeV scale and the mechanism underlying electroweak symmetry breaking. At the Large Hadron Collider, this process is statistically limited. Pushing the energy frontier beyond the LHC's reach will create new opportunities to exploit the rich phenomenology at higher centre-of-mass energies and luminosities. In this work, we perform a comparative analysis of the $hh+\text{jet}$ channel at a future 100 TeV hadron collider. We focus on the $hh\to b\bar b b\bar b$ and $hh \to b\bar b \tau^+\tau^-$ channels and employ a range of analysis techniques to estimate the sensitivity potential that can be gained by including this jet-associated Higgs pair production to the list of sensitive collider processes in such an environment. In particular, we observe that $hh \to b\bar b \tau^+\tau^-$ in the boosted regime exhibits a large sensitivity to the Higgs boson self-coupling and the Higgs self-coupling could be constrained at the 8\% level in this channel alone.}

\begin{document}

\date\today

\begin{flushright}
\footnotesize IPPP/18/10, LAPTH-003/18, CERN-TH-2018-023
\end{flushright}

\maketitle
\flushbottom
%%%%%%%%%%%%%%%%%%%%%%%%
\section{Introduction}
\label{sec:1}
The observed lack of any conclusive evidence for new interactions beyond the Standard Model (BSM) during the LHC's run-1 and the first 13 TeV analyses has tightly constrained a range of well-motivated BSM scenarios. For instance, the ATLAS and CMS collaborations have already set tight limits on top partners in supersymmetric~(e.g.~\cite{CMS-PAS-SUS-16-052,ATLAS-CONF-2017-034}) and strongly-interacting theories~(e.g.~\cite{ATLAS-CONF-2016-101,CMS-PAS-B2G-16-005}), which makes a natural interpretation of the TeV scale after the Higgs boson discovery more challenging than ever.

With traditional BSM paradigms facing increasing challenges as more data becomes available, a more bottom-up approach to parametrising potential new physics interactions has received attention recently. By interpreting Higgs analyses using Effective Field Theory (EFT), any heavy new physics scenario that is relevant for the Higgs sector can be investigated largely model-independently~\cite{Buchmuller:1985jz,Hagiwara:1986vm}, at the price of many ad hoc interactions to lowest order~\cite{Grzadkowski:2010es} in the EFT expansion.

Current measurements as well as first extrapolations of these approaches to the high luminosity (HL) phase of the LHC have provided first results as well as extrapolations of EFT parameters~\cite{Englert:2014uua,Englert:2015hrx,Falkowski:2015jaa,Corbett:2015ksa,Corbett:2015mqf}.
One of the parameters, which is particularly sensitive to electroweak symmetry breaking potential yet with poor LHC sensitivity prospects is the Higgs self-interaction. Constraining the trilinear self-interaction directly requires a measurement of (at least) $pp \to hh$~\cite{Dicus:1987ic,Glover:1987nx,Djouadi:1999rca,Plehn:1996wb,Baur:2002rb}; accessing quartic interactions in triple Higgs production is not possible at the LHC~\cite{Plehn:2005nk,Kling:2016lay} and seems challenging at future hadron colliders at best~\cite{Papaefstathiou:2015paa,Fuks:2017zkg}.
Early studies of the LHC's potential to observe  Higgs pair production have shown the most promising channels to be the $hh\to b\bar b \gamma\gamma$~\cite{Baur:2003gp} and $hh\to b\bar b \tau^+\tau^-$ channels~\cite{Baur:2003gpa,Dolan:2012rv}. Recent projections by ATLAS~\cite{ATL-PHYS-PUB-2017-001} and CMS~\cite{CMS:2015nat}, based on an integrated luminosity of 3 ab$^{-1}$ and on the pileup conditions foreseen for the HL-LHC, estimate a sensitivity to the di-Higgs signal in the range of
1-2$\sigma$. Recent phenomenological papers~\cite{Goertz:2013kp,Adhikary:2017jtu,Kim:2018uty}, combining the sensitivity to several different di-Higgs final states, reach similar conclusions. ATLAS~\cite{ATL-PHYS-PUB-2017-001} quotes a sensitivity to the value of the Higgs self-coupling (assuming SM-like coupling values for all other relevant interactions) in the range of  $-0.8 < \lambda/\lambda_{SM} < 7.7$,  at 95\% confidence limit.
Improving this sensitivity baseline is one of the main motivations of future high energy hadron colliders, and proof-of-principle analyses suggest that a vastly improved extraction of trilinear Higgs coupling should become possible~\cite{Yao:2013ika,Barr:2014sga,Azatov:2015oxa,He:2015spf,Contino:2016spe} at a future 100 TeV collider.

Most of these extrapolations have focused on gluon fusion production $p(g)p(g)\to hh$. Owing to large gluon densities at low momentum fractions, the associated di-Higgs cross section increases by a factor of $\sim39$ compared to 14 TeV collisions~\cite{deFlorian:2016spz,twiki}, with QCD corrections still dominated by additional unsuppressed initial state radiation~\cite{Dawson:1998py,deFlorian:2013jea,Borowka:2016ehy,Shao:2013bz,deFlorian:2015moa,Frederix:2014hta,Maltoni:2014eza}. While the process' kinematic characteristics of Higgs pair production remain qualitatively identical to the LHC environment, extra jet emission becomes significantly less suppressed leading to a cross section enhancement of $pp\to hhj$ of $\sim 80$\footnote{We impose $p_T(j) > 100$ GeV at the parton level.} compared to 14 TeV collisions. This provides another opportunity for the 100 TeV collider: Since the measurement of the self-coupling is largely an effect driven by the top quark threshold~\cite{Baur:2002rb}, accessing relatively low di-Higgs invariant masses is the driving force behind the self-coupling measurement. In fact, recoiling a collimated Higgs pair against a jet kinematically decorrelates $p_{T,h}$ and $m_{hh}$. Compared to $pp\to hh$, it thus exhibits a much higher sensitivity to the variation of the Higgs trilinear interaction while keeping $p_{T,h}$ large~\cite{Dolan:2012rv}, which is beneficial for the reconstruction and separation from backgrounds. However, such an approach is statistically limited at the LHC. Given the large increase in $pp\to hh+\text{jet}$ production in this kinematical regime as well as the increased luminosity expectations at a 100 TeV collider, it can be expected that jet-associated Higgs pair production can add significant sensitivity to self-coupling studies at a 100 TeV machine.

Quantifying this sensitivity gain in a range of exclusive final states with different phenomenological techniques is the purpose of this work. More specifically we consider final states with largest accessible branching fractions $hh \to b\bar b b\bar b$~\cite{Baur:2003gpa,deLima:2014dta,Behr:2015oqq} and $hh \to b\bar b \tau^+ \tau^-$~\cite{Baur:2003gpa,Dolan:2012rv,Barr:2013tda}, where we also differentiate between leptonic and hadronic $\tau$ decays (and consider their combination).

This work is organised as follows: We consider the $b\bar b \tau\tau$ channel in Sec.~\ref{sec:jbbtautau}. In particular we compare the performance gain of a fully-resolved di-Higgs final state analysis extended by substructure techniques highlighting the importance of high-transverse momentum Higgs pairs that are copious at 100 TeV. We discuss the $b\bar b b\bar b$ channel in Sec.~\ref{sec:jbbbb}.

%%%%%%%%%%%%%%%%%%%%%%%%%%%%
\section{The $jbb\tau\tau$ channels}
\label{sec:jbbtautau}
%%%%%%%%%%%%%%%%%%%%%%%%%%%%
%%%%%%%%%%%%%%%%%%%%%%%%%%%%
\subsection{General comments}
%%%%%%%%%%%%%%%%%%%%%%%%%%%%
Let us first turn to the $jbb\tau\tau$ channels. We will see that these are more sensitive to variations of the trilinear Higgs coupling and they therefore constitute the main result of this work. This is in line with similar studies at the LHC (see Refs.~\cite{Dolan:2012rv,deLima:2014dta,Barr:2013tda}) that show that the signal vs. background ratio can be expected to be better for this channel than for the four $b$ case.

We study the various decay modes of the taus and consider two exclusive final states, purely leptonic tau decays $h\to \tau_{\ell} \tau_{\ell}$ and mixed hadronic-leptonic decays
$h\to \tau_{\ell} \tau_h$, where the subscripts $\ell$ and $h$ denote the leptonic (to $e, \mu$) and hadronic decays of the taus, respectively. The scenario involving the purely hadronic decays, $h\to \tau_h \tau_h$ will undoubtedly add to the significance. However, scenarios involving two
hadronic taus will incur stronger QCD backgrounds and hence we will need to simulate various fake backgrounds and will also require an accurate knowledge of the $j \to \tau_j$ fake rate, where $j$ denotes a light jet. At this stage, we do not feel confident that we can reliably estimate these fake backgrounds, and
hence neglect this decay mode in the present study.

There are three categories of backgrounds that we consider for this scenario. The most dominant
background results from $t\bar{t}j$ with the leptonic top decays ($t \to b W \to b \ell \nu$), which includes decays to all the
three charged leptons~\footnote{The top decays to $b, \ell \; \textrm{and} \; \nu$ have been implemented as decay chains
in the \texttt{MadGraph5\_aMC@NLO} framework, at the production level.}. Furthermore, we have the pure EW background and a mixed QCD-EW
background of $j b \bar b \tau^+ \tau^-$~\footnote{The ``EW'' and ``QCD+EW'' processes correspond in
\texttt{MadGraph5\_aMC@NLO} to the interaction orders QED=4 QCD=1 (pure EW) and QED=2 QCD=3 (mixed QCD+EW)
respectively. We note that both classes of processes include single-Higgs production contributions.}.
The pure EW and QCD+EW processes consist of various sub-processes. A typical example for the
pure EW scenario is $pp\to HZ/\gamma^* + \; \textrm{jet} \to b\bar{b}\tau^+\tau^- + \; \textrm{jet}$. Whereas, for the QCD+EW processes, a
typical example is $pp\to b\bar{b} Z/\gamma^* + \; \textrm{jet} \to b\bar{b}\tau^+\tau^- + \; \textrm{jet}$. In all these background processes,
either from the $\tau$ decays or from the $W$-boson decays (for the $t\bar{t}j$ background), we may encounter leptons ($e,\mu$). 
There are potentially other irreducible backgrounds like $W \; (\to \ell \nu)+$ jets but these turn out to be
completely subdominant when compared to the other backgrounds. This is shown in the context of the $hh \to bb\tau\tau$ present and future analyses by ATLAS~\cite{ATL-PHYS-PUB-2015-046}
and CMS~\cite{CMS-PAS-HIG-16-012,Sirunyan:2017djm}. Similar conclusions will hold in the present study. Hence we neglect such backgrounds from our present analysis. All samples, including
the signal, are generated with \texttt{MadGraph5\_aMC@NLO}~\cite{Alwall:2014hca} in Born-level mode, and we neglect effects from jet merging up to higher jet multiplicities. For our signal samples, the Higgs bosons are decayed using \texttt{MadSpin}~\cite{Frixione:2007zp,Artoisenet:2012st}; the showering is performed using \texttt{Pythia 8}~\cite{Sjostrand:2014zea}. To account for QCD corrections we use global $K$ factors for the signal of $K=1.8$ for the EW contributions (extrapolating from~\cite{Binoth:2009wk}), $K=1.5$ for the QCD+EW contribution~\cite{Campbell:2002tg} as well as $K=1.0$ for $t\bar t j$ following~\cite{Bevilacqua:2015qha}.

To operate with an efficient Monte Carlo tool chain, we generate the EW and mixed QCD+EW events with the following generator level cuts: $p_T^{b} > 23$ GeV, $p_T^{\ell}> 8$ GeV, $|\eta^{b,\ell}| < 3$, $p_T^j > 100$ GeV, $|\eta^j| < 5$, $\Delta R_{b,b}>0.2$, $\Delta R_{\ell \ell} > 0.15$, $\Delta R_{b/j,\ell}>0.3$, $90~\text{GeV} < M_{b,b} < 160~\text{GeV}$ and $90~\text{GeV} < M_{\ell, \ell} < 200~\text{GeV}$, where $\ell = e, \mu, \tau$ and $b$ denotes final state bottom quarks. $R$ is the azimuthal angle---pseudo-rapidity ($\phi$-$\eta$) distance and $M$ denotes invariant masses. The same requirements are imposed on $t\bar t j$, however, without a lower bound on $M_{\ell\ell}$. The only event generator cut applied to the signal is transverse momentum cut on the light flavor jet $p_T^j > 100$ GeV.

Given the discriminating power of $m_{T2}$ which was motivated in Ref.~\cite{Barr:2013tda}  to reduce the $t\bar t$ background, we consider a similar variable with the aim to reduce the dominant $t\bar{t}+{\text{jet}}$ background. The top background final state can be described schematically through a decay chain
\begin{subequations}
\label{eq:mt2dec}
\begin{eqnarray}
\label{eq:unpr}
 A &\to & B + C\\
 A &\to & B' + C',
\label{eq:pr}
\end{eqnarray}
where $B,B'$ ($C,C'$) denote the visible (invisible) decay products of the top branching ($A=t,\bar t$). For such a branching one can construct the $m_{T2}$ variable~\cite{Lester:1999tx}
\begin{equation}
\label{eq:mt2}
 m_{T2}(m_B,m_{B'},\mathbf{b}_T,\mathbf{b}'_T,\mathbf{p}_T^{\Sigma},m_C,m_{C'}) \equiv \underset{\mathbf{c}_T + \mathbf{c}'_T =
 \mathbf{p}_T^{\Sigma}} {\text{min}} {\text{max}(m_T,{m}'_T)},
\end{equation}
where $m_T$ denotes the transverse mass constructed from $\mathbf{b}_T$, $\mathbf{c}_T$
and $m_B$
\begin{equation}
 m_T^2(\mathbf{b}_T, \mathbf{c}_T, m_B, m_C) \equiv m_B^2 + m_C^2 + 2(e_B e_C - \mathbf{b}_T \cdot \mathbf{c}_T),
\end{equation}
with transverse energy $e_i^2 = m_i^2 + \mathbf{p}_{i,T}^2$, $i=B,C$.
$m'_T$ refers to the same observable calculated from the primed quantities in Eq.~\eqref{eq:mt2dec}. The minimisation in Eq.~\eqref{eq:mt2} is performed over all momenta $\mathbf{c}_T$ and $\mathbf{c}'_T$, subject to the condition that their sum needs to reproduce the correct $\mathbf{p}_T^{\Sigma}$, which is normally chosen to coincide with the overall missing energy $\vec{\cancel{p}}_T$. However, because the tau's decay is partially observable, we can modify the $m_{T2}$ definition to include the visible transverse momenta of the tau leptons by identifying
\begin{equation}
 \mathbf{p}_T^{\Sigma} \equiv \vec{\cancel{p}}_T + \mathbf{p}_T^{\text{(vis)}}(\tau) + \mathbf{p}_T^{\text{(vis)}}(\tau')
                      =      \mathbf{p}_T(W) + \mathbf{p}_T(W').
\end{equation}
\end{subequations}
As we will see below, this modified $m_{T2}$ plays a crucial role in suppressing the dominant $t\bar{t}j$ background.
We must emphasise here that many distinctly different definitions of $m_{T2}$ have been considered in Ref.~\cite{Baringer:2011nh}.
The authors in Ref.~\cite{Barr:2013tda} have considered several such definitions of the $m_{T2}$ variable and found
them having very similar discriminatory power.

%%%%%%%%%%%%%%%%%
\subsection{The resolved $\tau_{\ell} \tau_{\ell}$ channel}
\label{sec:2.2.1}
%%%%%%%%%%%%%%%%%
The leptonic di-tau final states are undoubtedly the cleanest channels out of the three di-tau options. We can identify exactly two leptons ($e, \mu$), two $b$-tagged jets and
at least one hard non $b$-tagged jet.  We therefore pre-select the events by requiring the following cuts at
reconstruction level\footnote{Jets are defined through the anti-$kT$ algorithm~\cite{Cacciari:2008gp,Cacciari:2011ma} with a jet resolution parameter $0.4$ inside the rapidity range $|\eta^j| < 4.5$. For the $b$-tagging efficiency we choose 60\% at a 2\% mistagging rate, which is a realistic at the LHC~\cite{ATLAS:2012ima}. Isolated muons and electrons are defined by requiring a small hadronic energy deposit in the vicinity of the lepton candidate, $E^{\text{had}}/p_T^{\ell}<10\%$ within $\Delta R<0.2$.}: jets are clustered with size 0.4 and $p_T^{j}>30$~GeV in $|\eta|<4.5$; the hardest jet is required to have $p_T^{j_1}>105~\text{GeV}$. Leptons are required to have $p_T^{\ell}>10~\text{GeV}$ and $|\eta| < 2.5$. We require two leptons and select two jets with $p_T > 30$ GeV and $|\eta| < 2.5$, which are subsequently $b$-tagged. All objects need to be well separated $\Delta R(b,b/j_1/\ell),\Delta R(\ell,j_1) >0.4$ and $\Delta R(\ell,\ell)>0.2$. To efficiently suppress the $Z$-induced background we demand  $105~\text{GeV}  < M_{b,b}  < 145 ~\text{GeV}$. Furthermore we require a significant amount of missing energy $\cancel{E}_T > 50$ GeV.
%imposed in order to reduce sub-leading backgrounds of the form $j b b \ell \ell \nu \nu$ which we have not simulated as part of our backgrounds.

After these pre-selection requirements we apply a boosted decision tree (BDT) analysis which is the experiments' weapon of choice when facing a small signal vs. background ratio (see e.g. the very recent ATLAS $t\bar t h$ analysis~\cite{Aaboud:2017rss}). We include a large amount of (redundant) kinematic information~\footnote{The redundant variables which do not affect the significance are:
$p_T^{b_2}, p_T^{\ell_2}, \Delta R (b_2 \ell_2), \Delta R (\ell_2 j)$ and $\Delta \phi (bb,\ell\ell)$, where 
the index refers to the $p_T$ ordering of an object.} to the training phase, as listed in Tab.~\ref{tab:1}.\footnote{We have checked our results for overtraining.}

%%%%%%%%%%%%%%%%%
\begin{table}[!t]
\centering
\parbox{0.45\textwidth}{
\begin{tabular}{| c  | l |}
\hline
observable & reconstructed object \\
\hline
\multirow{5}{*}{$p_T$} & 2 $b$-tagged jets\\
& 2 leptons \\
& hardest non $b$-tagged jet\\
& $bb$ system \\
& $\ell \ell$ system\\
\hline
\multirow{2}{*}{$p_T$ ratios} & 2 $b$-tagged jets \\
& 2 leptons \\
\hline
\multirow{4}{*}{$\Delta R$} & 2 $b$-tagged jets \\
& 2 leptons \\
& $b$-tagged jets and jet $j_1$/leptons\\
& leptons and jet $j_1$\\
\hline
\multirow{3}{*}{$M$} & 2 $b$-tagged jets \\
& 2 leptons \\
& $b$-tagged jets and leptons\\
\hline
$m_{T2}$ & described in Eq.~\eqref{eq:mt2dec} \\
\hline
$\Delta \phi$  & between $bb$ and $\ell \ell$ systems \\
\hline
$\cancel{E}_T$ &  reduce sub-leading backgrounds \\
\hline
\end{tabular}
}
\hfill\parbox{0.4\textwidth}{
\vspace{8cm}
\caption{\label{tab:1} Observables included in the boosted decision tree for the leptonic $\tau$ channels of the $pp\to hhj$ analysis of Sec.~\ref{sec:2.2.1}.}
}
\end{table}
%%%%%%%%%%%%%%%%%

We focus on a training of the boosted decision tree for a SM-like value of the trilinear Higgs coupling $\lambda_{\text{SM}}$. We employ the boosted decision tree algorithm of the \texttt{TMVA} framework~\cite{Hocker:2007ht} on the basis of 30~$\text{ab}^{-1}$ of data at 100 TeV. Our results are tabulated in Tab.~\ref{tab:2}. As can be seen, we can typically expect small signal vs background ratios at small signal cross sections. The latter is mostly due to the small fully-leptonic branching ratios of the tau pairs.

%%%%%%%%%%%%%%%%%%%%%%%%%%%%%%%
\begin{table}[ht]
\centering
\begin{footnotesize}
\begin{tabular}{| c  c | c | c | c | c || c | c |}
\hline
& signal  & QCD+EW  & EW  & $t\bar t j$  & tot. background & $S/B$ & $S/\sqrt{B}$, 30/ab \\
\hline
$\kappa_\lambda=1/2$ & 0.070 & \multirow{3}{*}{0.26} &  \multirow{3}{*}{0.04} &   \multirow{3}{*}{1.25}  &  \multirow{3}{*}{1.55} & 0.046 & 9.85 \\
$\kappa_\lambda=1$ & 0.059 & & & & & 0.038 & 8.19 \\
$\kappa_\lambda=2$ & 0.043 & & & & & 0.028 & 5.98 \\
\hline
\end{tabular}
\end{footnotesize}
\caption{\label{tab:2} Results of the fully leptonic tau decay channels outlined in Sec.~\ref{sec:2.2.1} in femtobarns (numbers to the left of the double vertical lines) after an optimised cut on the BDT output. We include results for three different choices of the self-coupling within the $\kappa$ framework~\cite{Heinemeyer:2013tqa}, $\kappa_\lambda=\lambda/\lambda_{\text{SM}}$; BDT training is performed with $\lambda=\lambda_{\text{SM}}$.}
\end{table}
%%%%%%%%%%%%%%%%%%%%%%%%%%%%%%%

%%%%%%%%%%%%%%%%%
\subsection{The resolved $\tau_{\ell} \tau_{h}$ channel}
\label{sec:2.2.2}
%%%%%%%%%%%%%%%%%
Given the small $S/B$ for the fully leptonic channel of the previous section we consider the case where one tau lepton decays leptonically while the other tau decays hadronically.

Recently, a major CMS level-1 trigger update has increased the hadronic tau tagging efficiency by a factor of two~\cite{Cadamuro:2015lbd,CMS-DP-2015-009,Mastrolorenzo:2016dyo} for tau candidates with $p_T\gtrsim 20~\text{GeV}$, robust against pile-up effects. Fully-hadronic di-tau decays of the Higgs boson for 13 TeV collisions can be tagged at 70\% with a background rejection of around 0.999. These improvements suggest that a single tau tagging performance of 70\% in a busier environment of the $hhj$ final state at 100~TeV is not unrealistic and we adopt this working point in the following, assuming a sufficiently large background rejection for fakes to be negligible.

\begin{figure}[ht]
\centering
\includegraphics[scale=0.35]{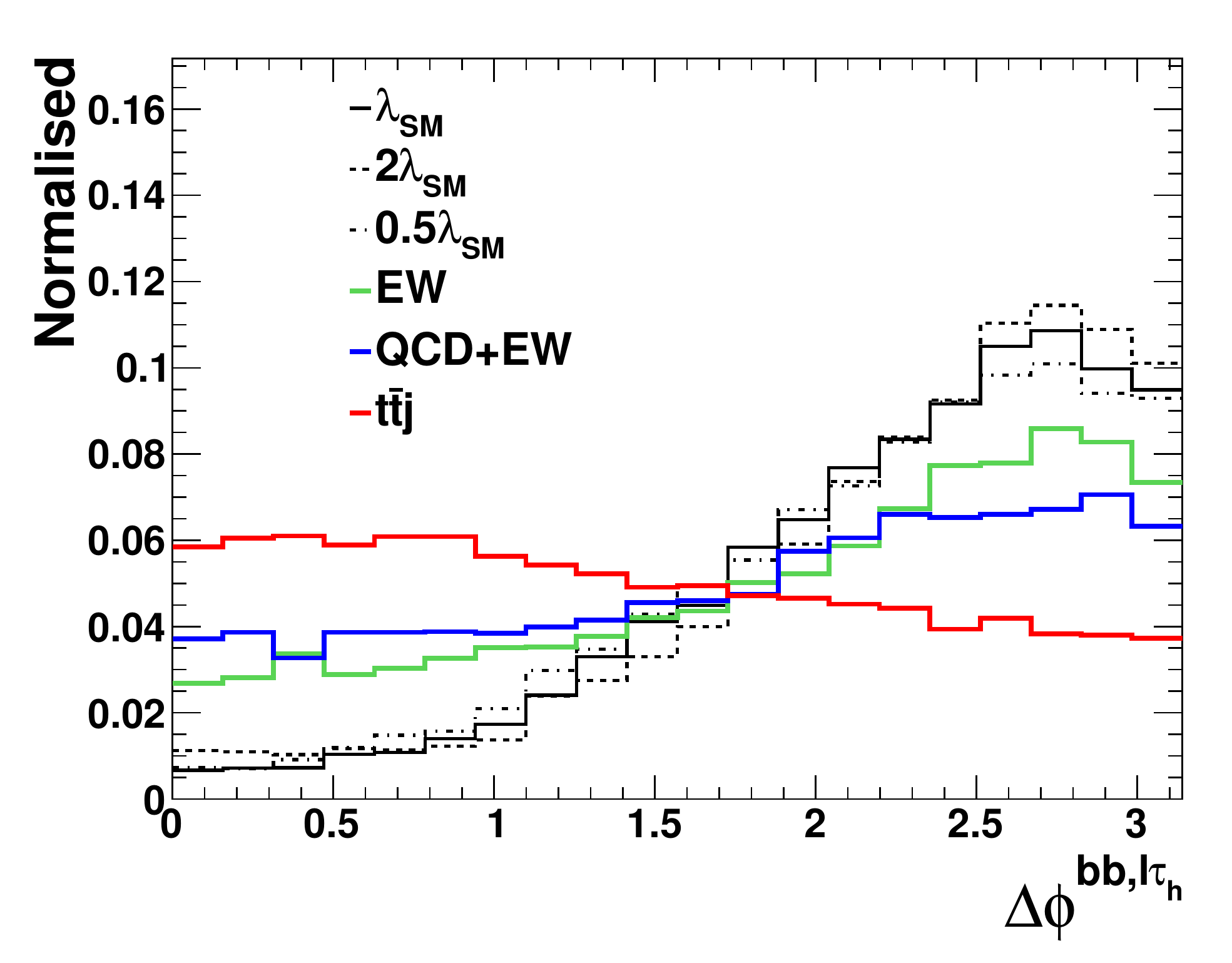}\hfill \includegraphics[scale=0.35]{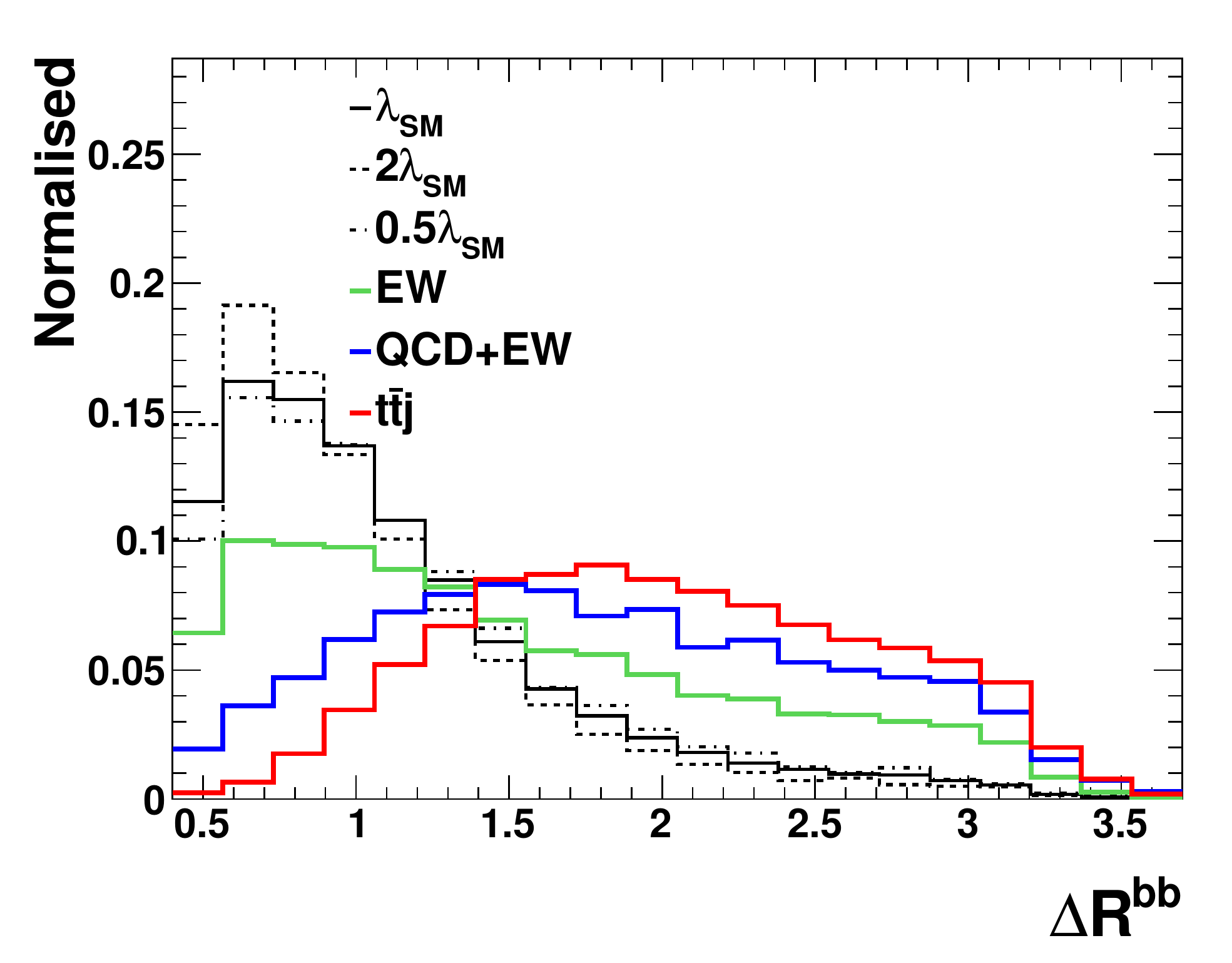}\\
\includegraphics[scale=0.35]{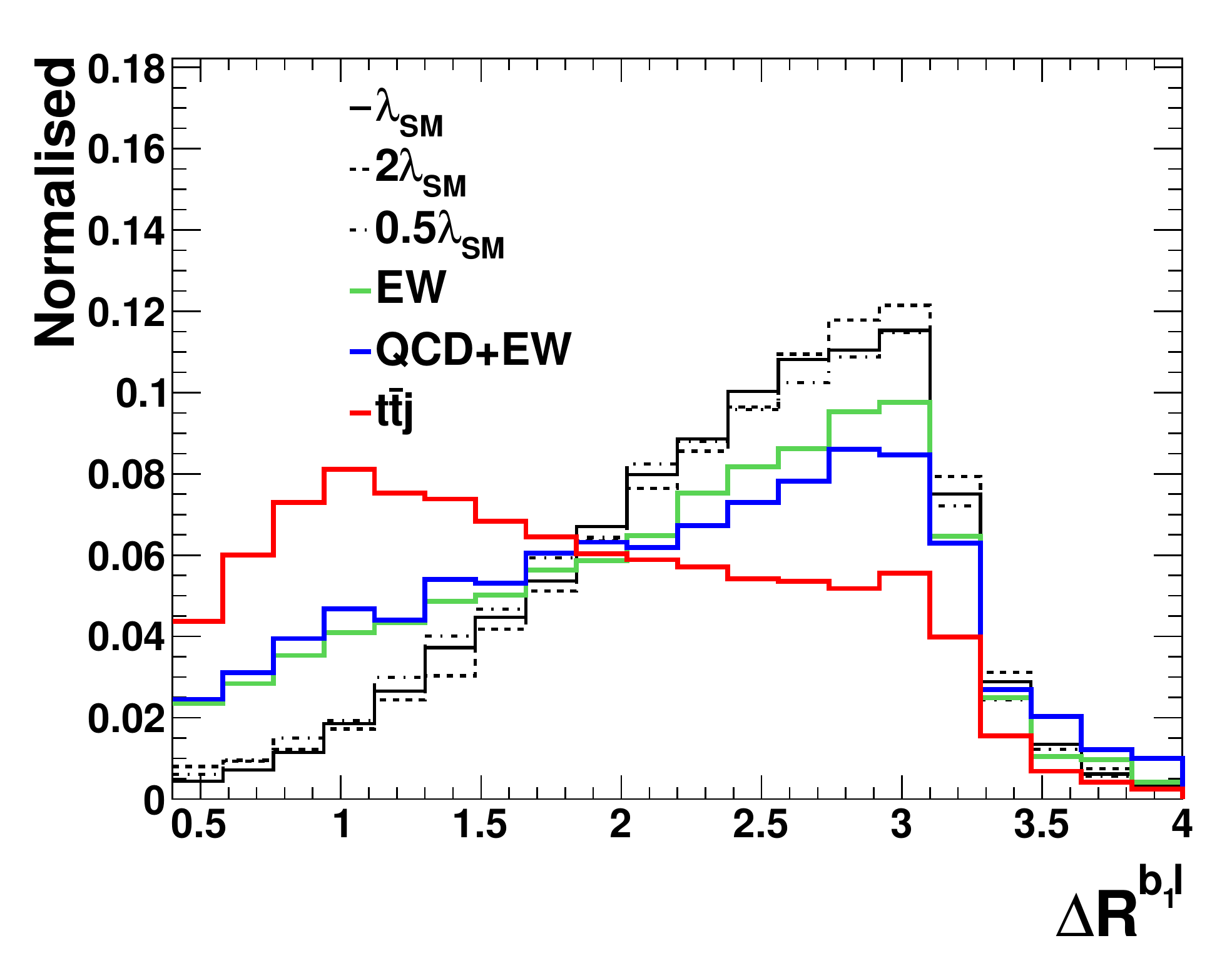}\hfill \includegraphics[scale=0.35]{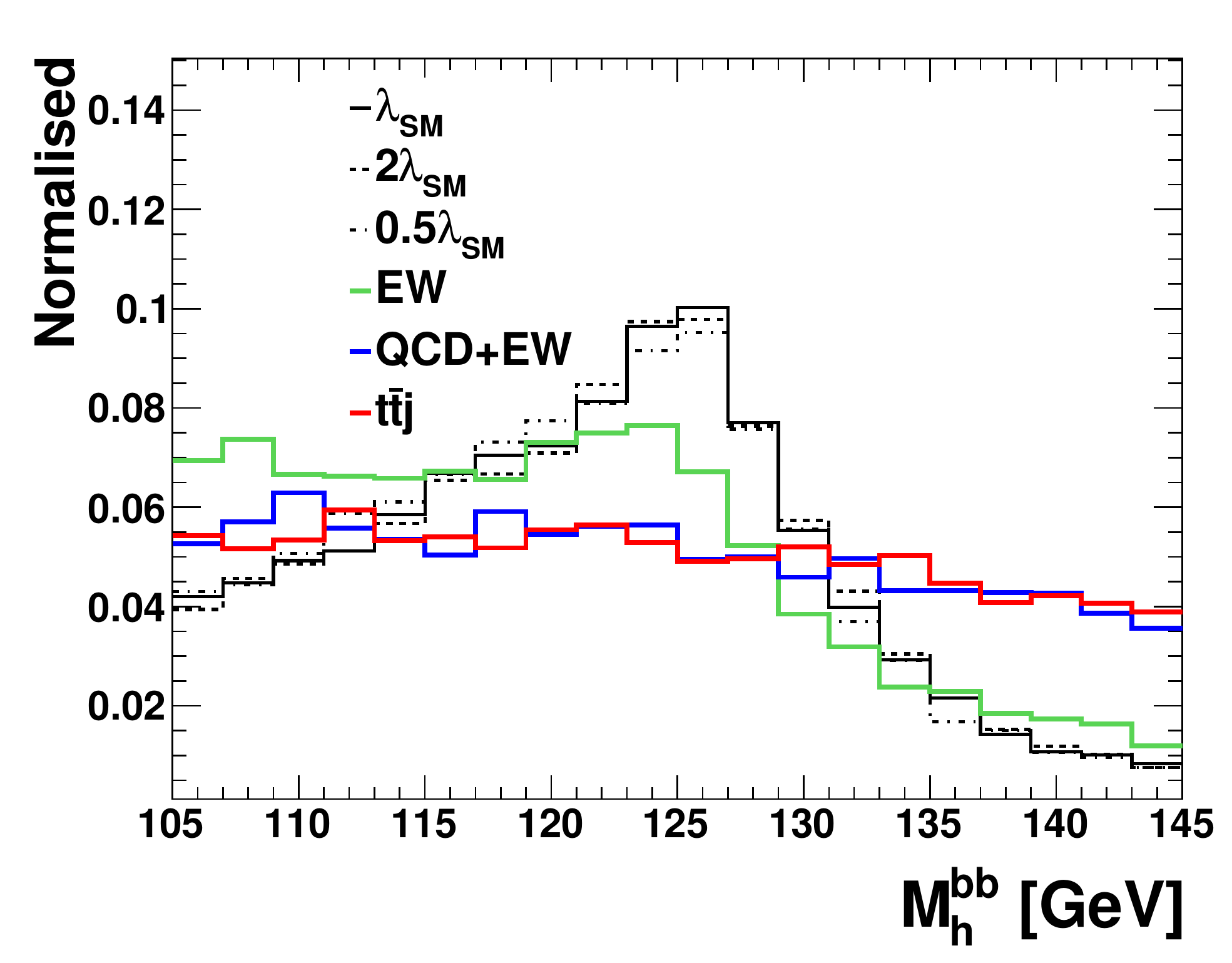}
\caption{Normalised differential distributions that serve to isolate signal from background in the $\tau_{\ell} \tau_h$ case in the BDT analysis.}
\label{fig:taLtaH_resolved1}
\end{figure}

\begin{figure}[ht]
\includegraphics[scale=0.35]{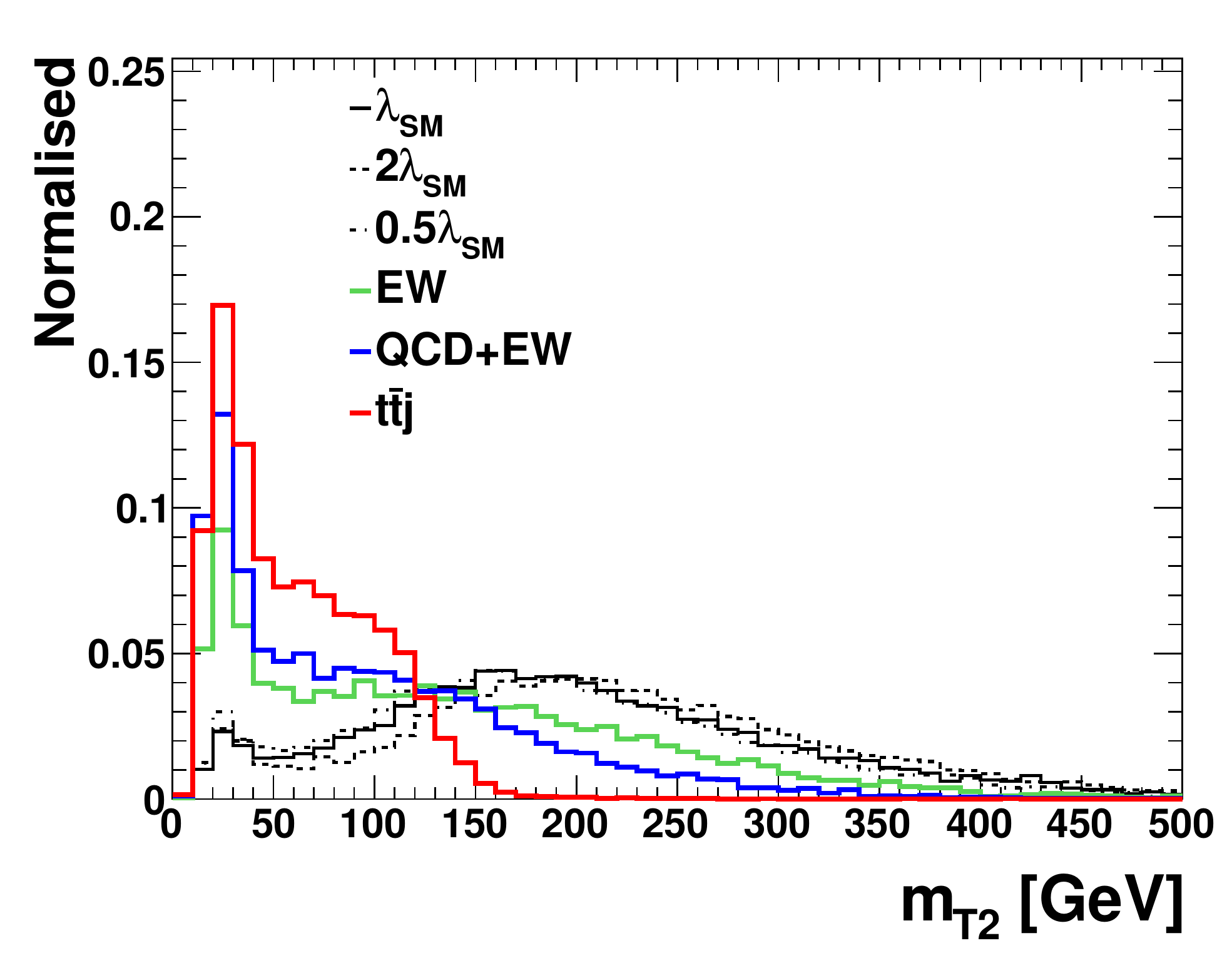}\hfill\includegraphics[scale=0.35]{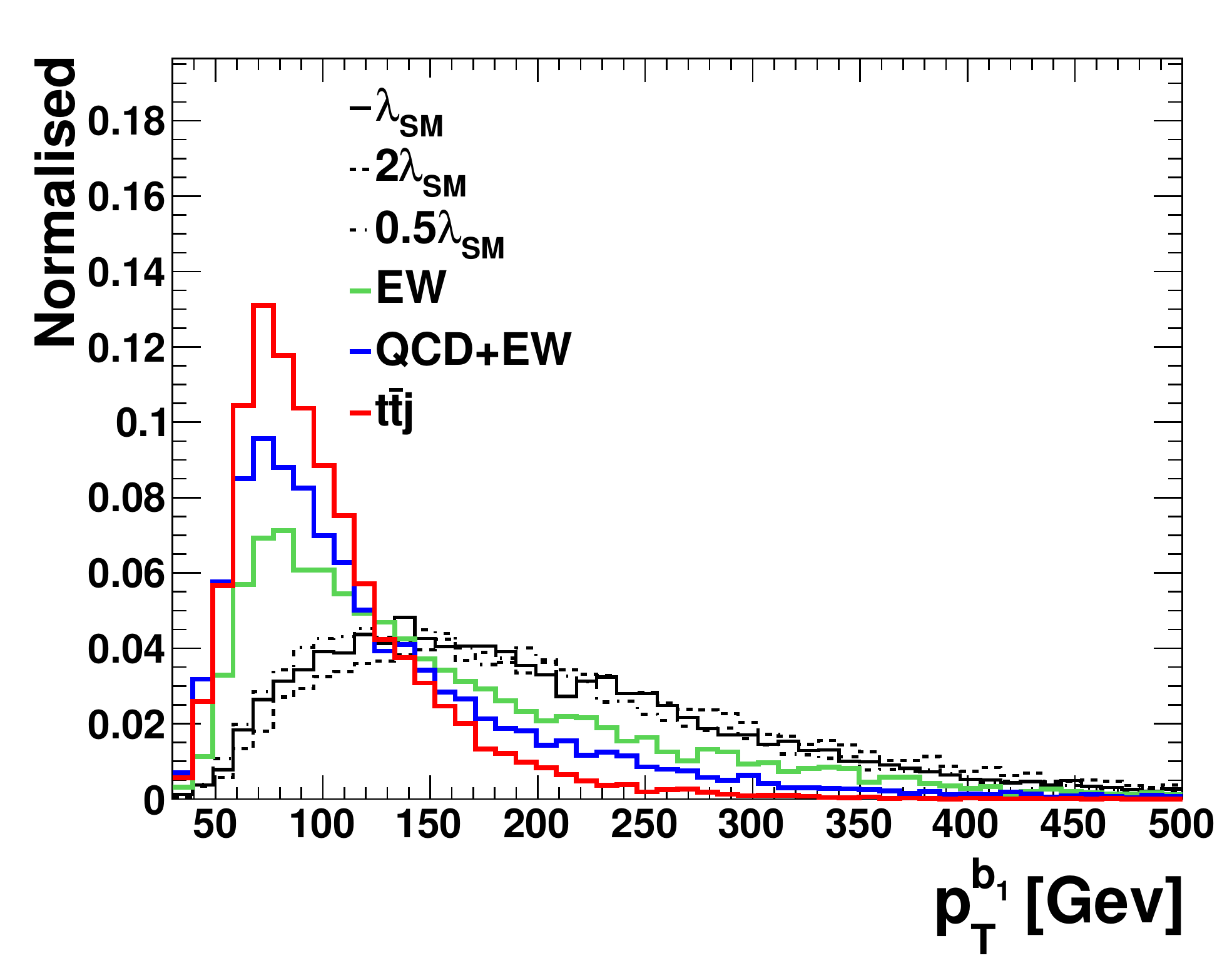}\\
\includegraphics[scale=0.35]{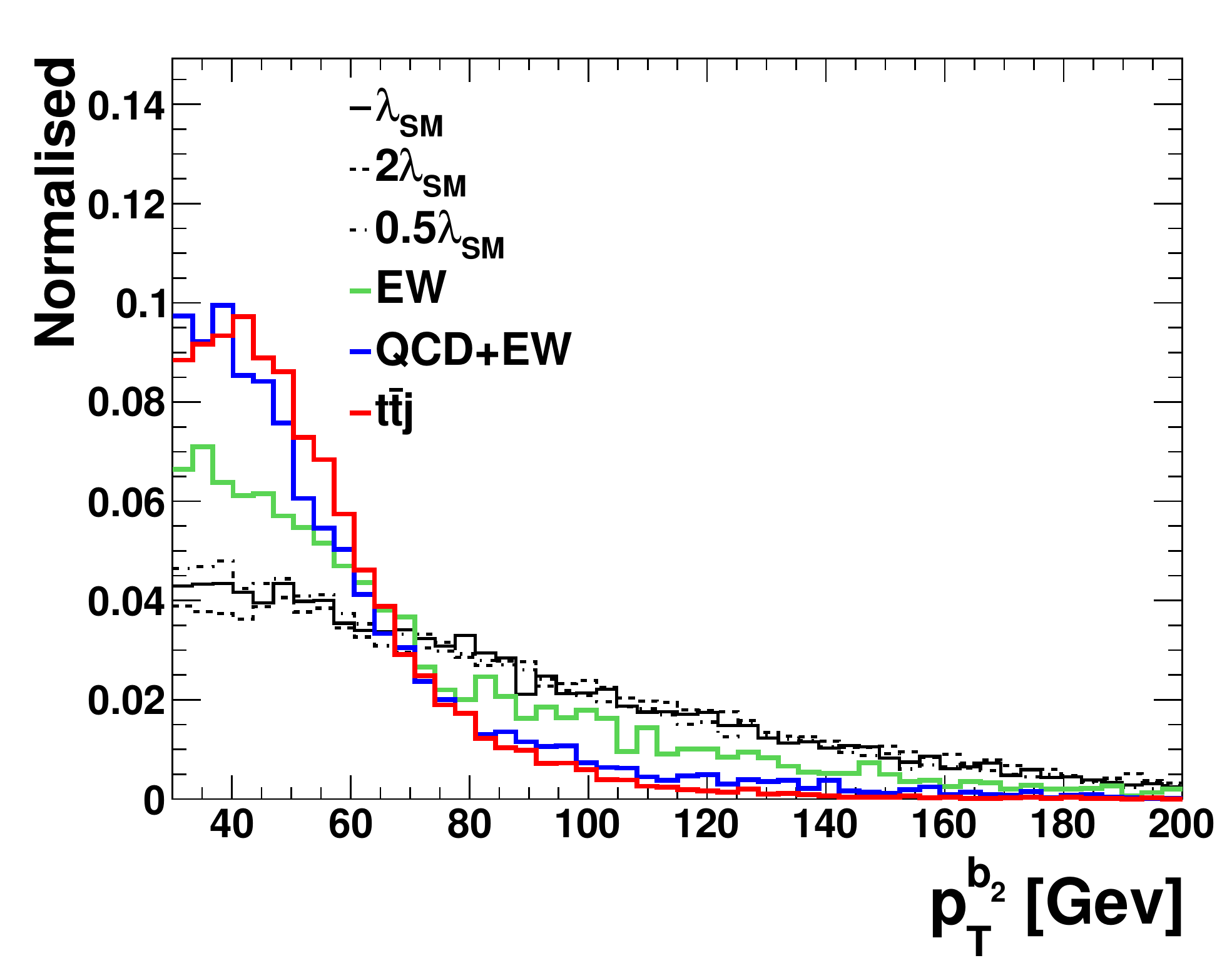}\hfill\includegraphics[scale=0.35]{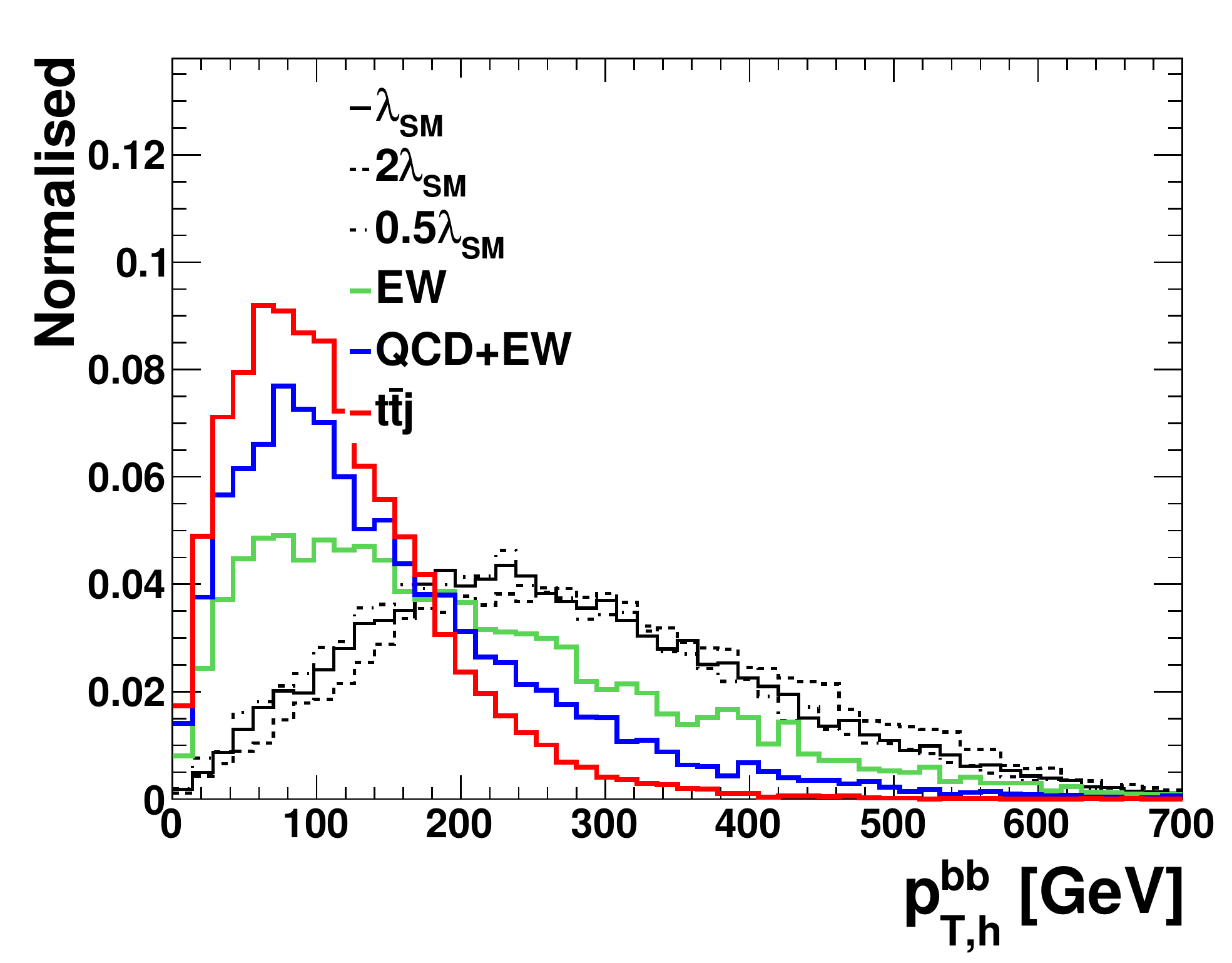}
\caption{Normalised differential distributions that serve to isolate signal from background in the $\tau_{\ell} \tau_h$ case in the BDT analysis.}
\label{fig:taLtaH_resolved2}
\end{figure}

\begin{figure}[ht]
\centering
\includegraphics[scale=0.35]{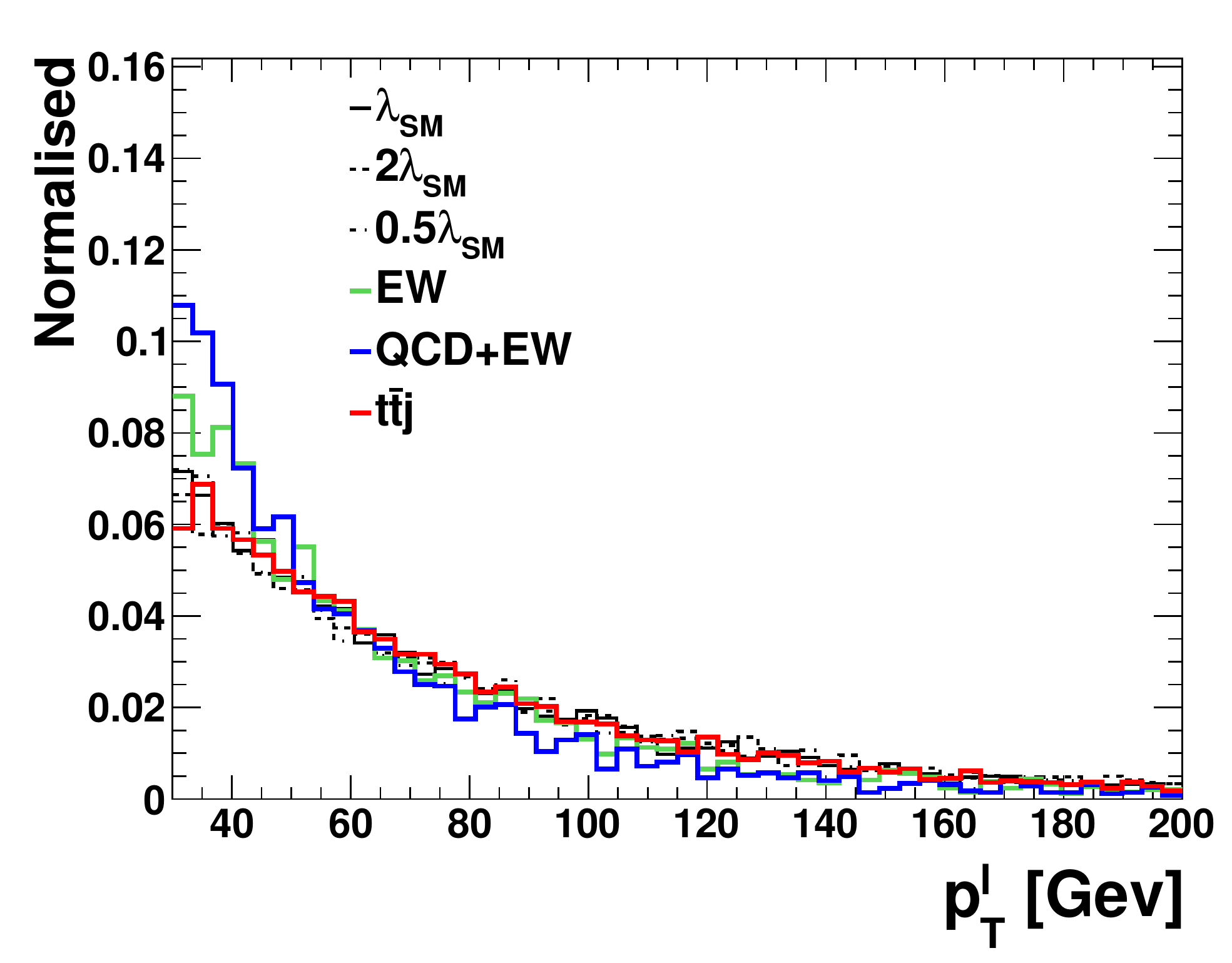}\hfill\includegraphics[scale=0.35]{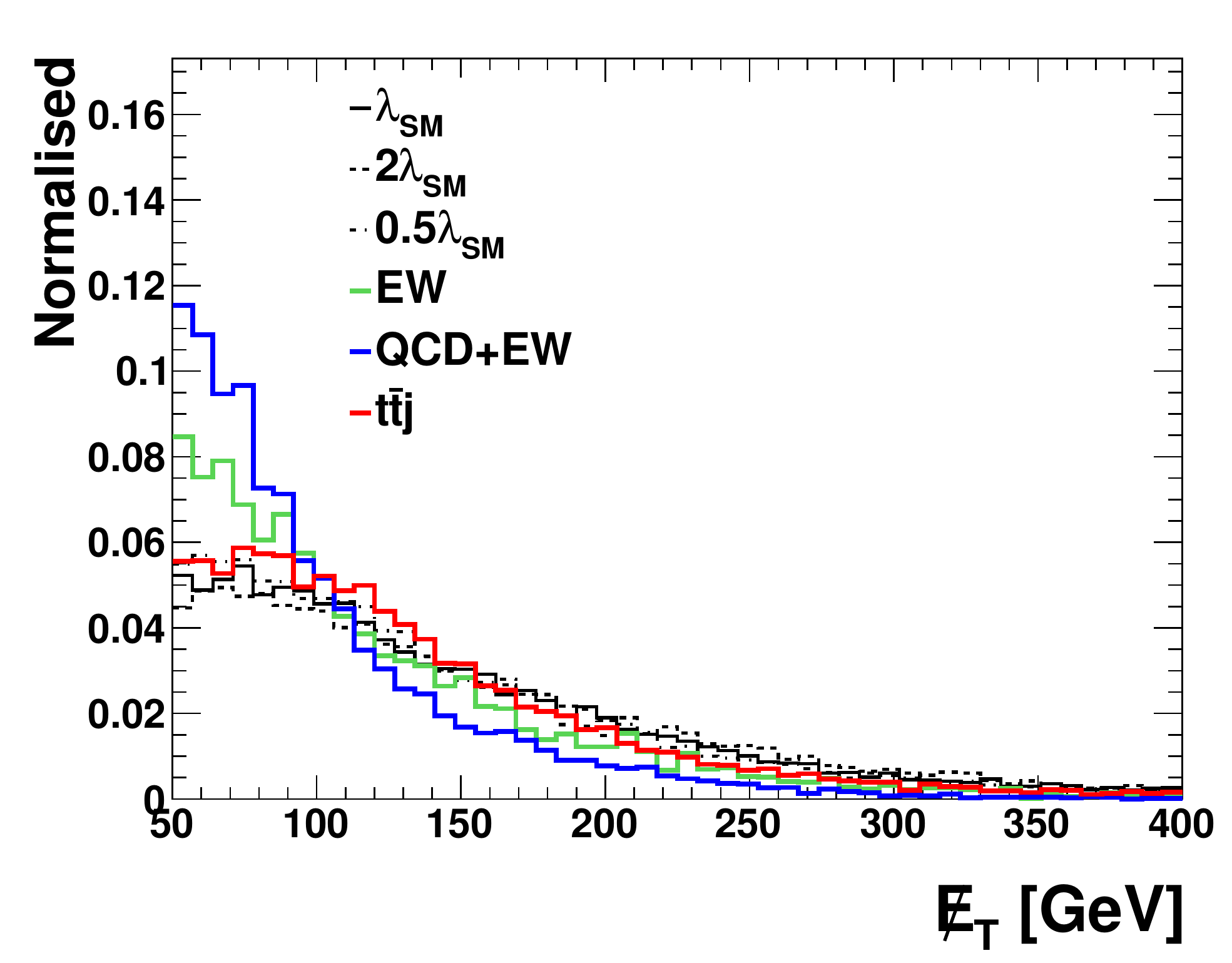}\\
\parbox{0.45\textwidth}{\includegraphics[scale=0.35]{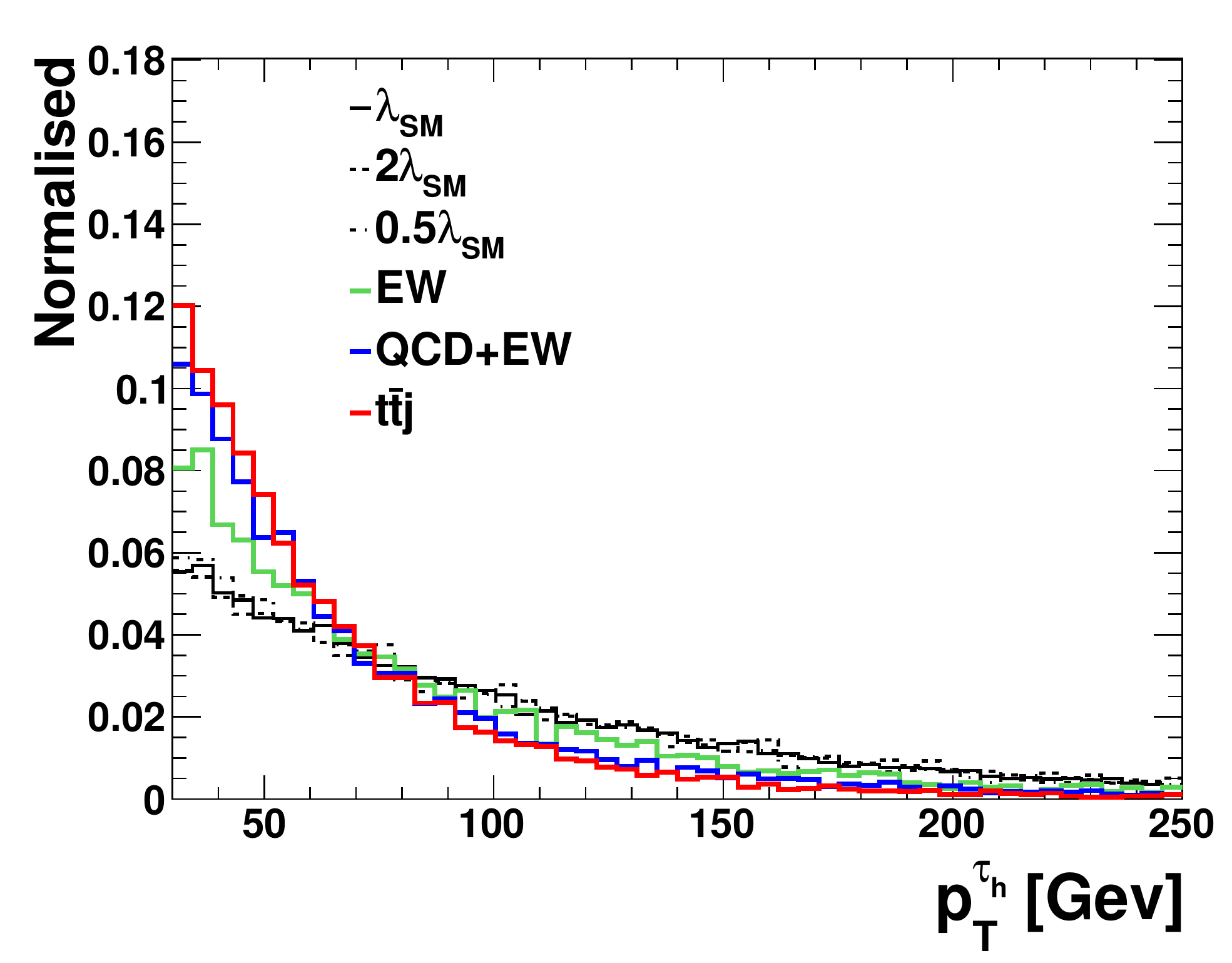}}
\hfill
\parbox{0.45\textwidth}
{\caption{\label{fig:taLtaH_resolved3} Normalised differential transverse momentum contributing to the $\tau_{\ell} \tau_h$ BDT analysis.}}
\end{figure}

We follow the analysis of the previous section and employ similar variables for the BDT. The only difference here is that here we demand 2 $b$-tagged jets, one
$\tau$-tagged jet, one lepton and at least one hard non $b,\tau$-tagged jet. All the aforementioned variables for the $\tau_{\ell}
\tau_{\ell}$ scenario, Tab.~\ref{tab:1}, can be utilised here with the only difference of replacing one lepton by a $\tau_h$~\footnote{The redundant 
variables for this case are: $\Delta R (b_2 \ell), \Delta R (b_2 \tau_h), \Delta R (b_1 j), \Delta R (b_2 j)$ 
and $\Delta R (\tau_h j)$.}. The distributions are shown in Figs.~\ref{fig:taLtaH_resolved1}-\ref{fig:taLtaH_resolved3} and the results are tabulated in Tab.~\ref{tab:3}. As can be seen, different to the fully-leptonic case, the increase in signal allows us to suppress the dominant $t\bar tj$ background further without compromising the signal count too much. This leads to a much larger expected sensitivity in the $\tau_\ell\tau_h$ channels.

%%%%%%%%%%%%%%%%%%%%%%%%%%%%%%%

%%%%%%%%%%%%%%%%%%%%%%%%%%%%%%%

%%%%%%%%%%%%%%%%%%%%%%%%%%%%%%%
%\begin{figure}[ht]
%\centering
%\includegraphics[scale=0.35]{Figures/Resolved/ROC_lam1.pdf}
%\caption{BDT output observable, \checkthis{details}.}
%\label{fig:taLtaH_substructure4}
%\end{figure}
%%%%%%%%%%%%%%%%%%%%%%%%%%%%%%%

%%%%%%%%%%%%%%%%%%%%%%%%%%%%%%%
\begin{table}[ht]
\centering
\begin{footnotesize}
\begin{tabular}{| c  c | c | c | c | c || c | c |}
\hline
& signal  & QCD+EW  & EW  & $t\bar t j$  & tot. background & $S/B$ & $S/\sqrt{B}$, 30/ab \\
\hline
$\kappa_\lambda=0.5$ & 0.169 & \multirow{3}{*}{0.52} &  \multirow{3}{*}{0.07} &   \multirow{3}{*}{0.37}  &  \multirow{3}{*}{0.96} &  0.176 & 29.81 \\
$\kappa_\lambda=1$ & 0.141 & & & & & 0.147 & 24.97 \\
$\kappa_\lambda=2$ & 0.105 & & & & & 0.109 & 18.49 \\
\hline
\end{tabular}
\end{footnotesize}
\caption{\label{tab:3} Results of the $h\to \tau_\ell\tau_h$ decay channels outlined in Sec.~\ref{sec:2.2.2} in femtobarns (numbers to the left of the double vertical lines) after an optimised cut on the BDT output. We include results for three different choices of the self-coupling within the $\kappa$ framework~\cite{Heinemeyer:2013tqa}, $\kappa_\lambda=\lambda/\lambda_{\text{SM}}$; BDT training is performed with $\lambda=\lambda_{\text{SM}}$.}
\end{table}
%%%%%%%%%%%%%%%%%%%%%%%%%%%%%%%

Combining the results of the previous section with the $\tau_\ell \tau_h$ results into a log-likelihood CLs hypothesis test~\cite{Junk:1999kv,Read:2000ru,Read:2002hq} assuming the SM as null hypothesis values of (assuming no systematic uncertainties)

\begin{align}
\label{eq:resolvedresult}
0.65 < \kappa_\lambda < 1.44 \quad & 3/\text{ab}\,, \\
0.88 < \kappa_\lambda < 1.13 \quad & 30/\text{ab}\,,
\end{align}
at 68\% confidence level. Here, $\kappa_{\lambda} = \lambda/\lambda_{\textrm{SM}}$, is the measure of the
deviation of the Higgs trilinear coupling with respect to the SM expectation.

\subsection{The significance of high-$p_T$ final states}
\label{sec:substruc}
So far our strategy has focused on resolved particle-level objects without making concessions for the larger expected sensitivity of the high $p_T$ final states. Jet-substructure techniques~(see e.g.~\cite{Butterworth:2008iy}) are expected to be particularly suited for kinematic configurations for which $h\to b\bar b$ recoils against the light-flavor and hard jet~\cite{Dolan:2012rv}, while the $h\to \tau \tau$ decay happens at reasonably low transverse momentum. This way, although one Higgs is hard, low invariant Higgs pair-masses can be accessed from an isotropic $h\to \tau \tau$ decay given a collimated $b\bar b$ pair. This particular kinematic configuration is not highlighted in the previous section and we can expect that the sensitivity of Eq.~\eqref{eq:resolvedresult} will increase once we focus with jet-substructure variables on this phase-space region which is highly relevant for our purposes. The benefit of this analysis will hence be two-fold: firstly we will exploit the background rejection of the non-Higgs final states through the adapted strategies of jet-substructure techniques. And secondly we will directly focus on a phase space region where we can expect the impact of $\kappa_\lambda\neq 1$ to be most pronounced.

%%%%%%%%%%%%%%%%%%%%%%%%%%%%%%%
\begin{figure}[ht]
\includegraphics[scale=0.35]{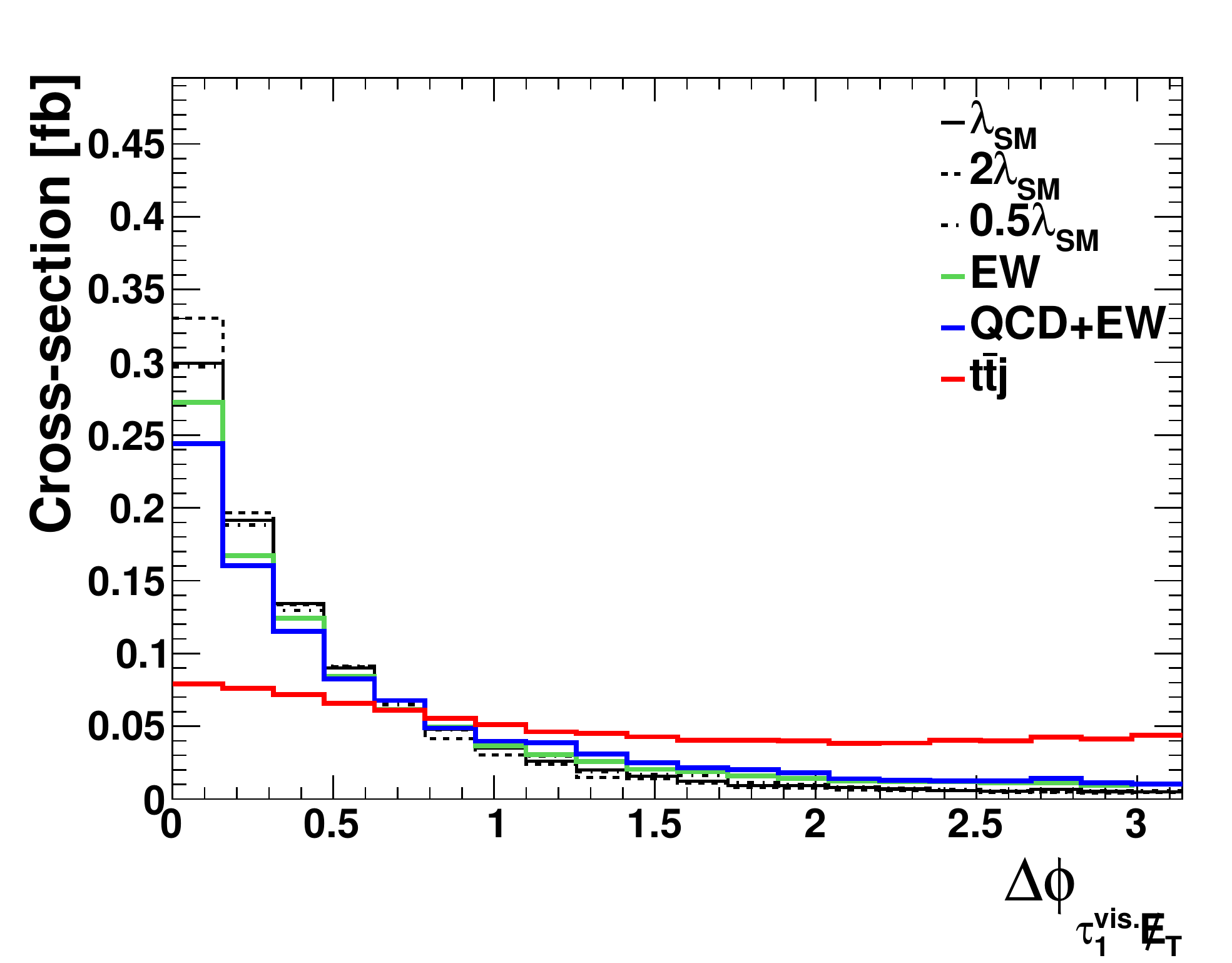}\hfill \includegraphics[scale=0.35]{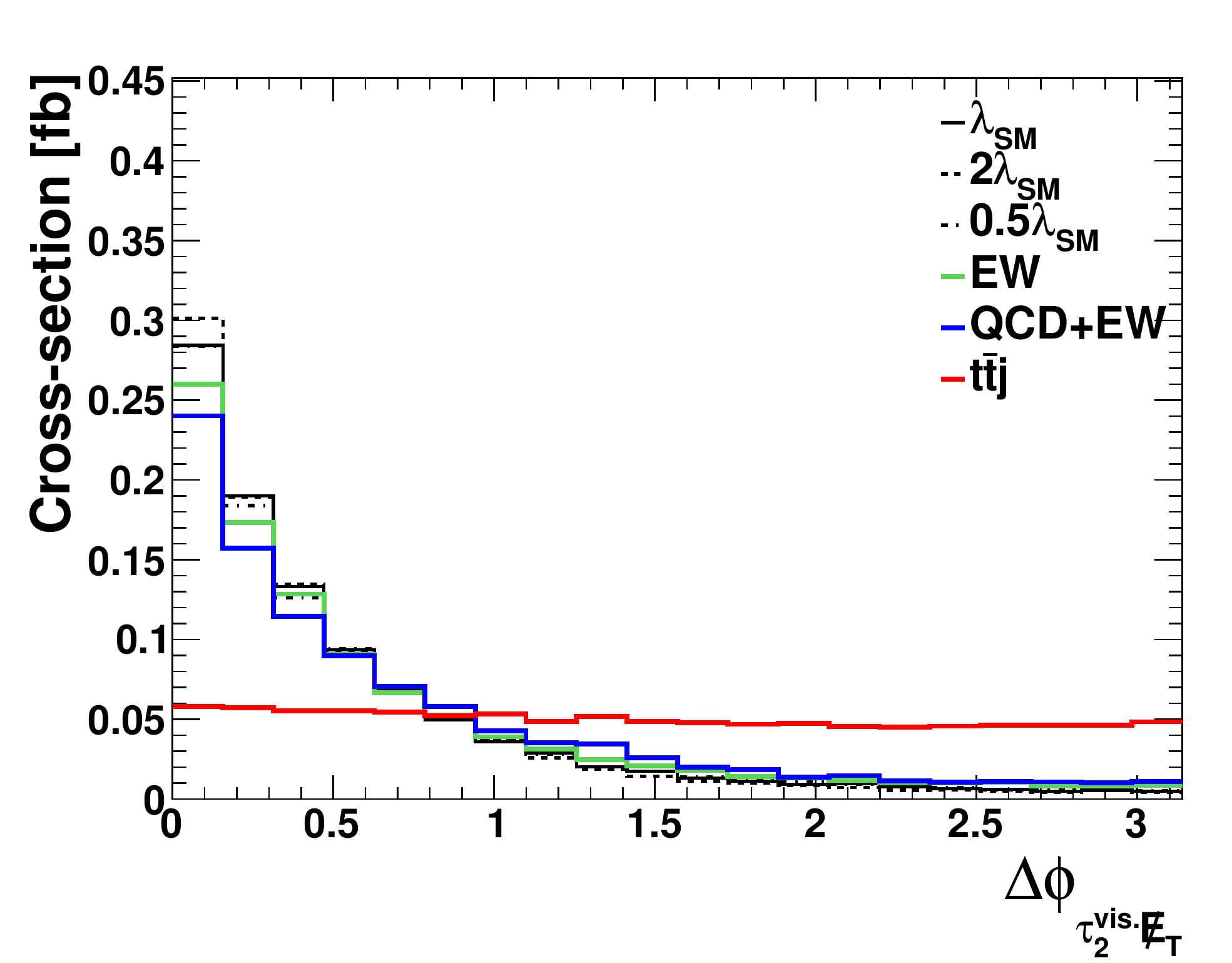}\\
\includegraphics[scale=0.35]{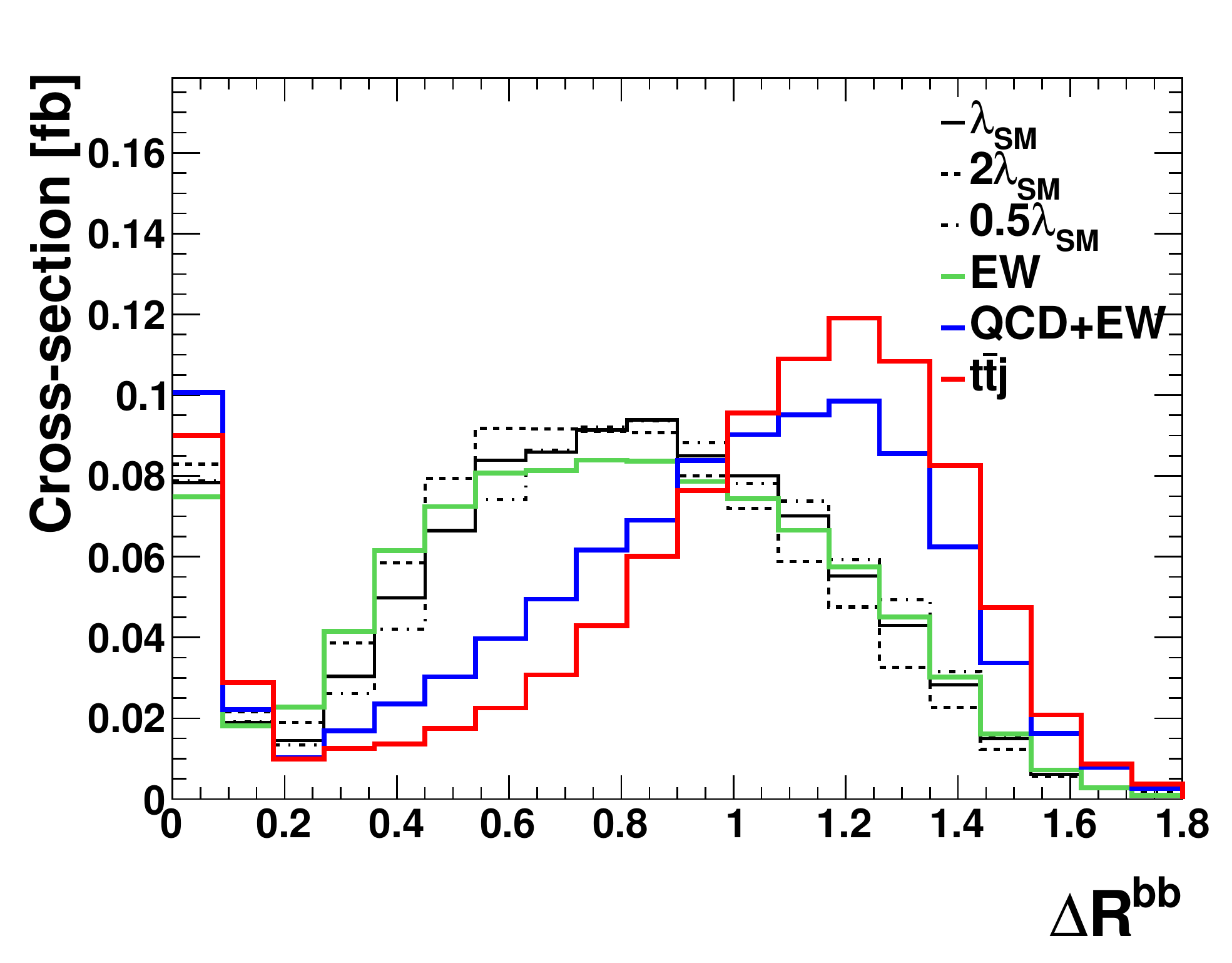}\hfill \includegraphics[scale=0.35]{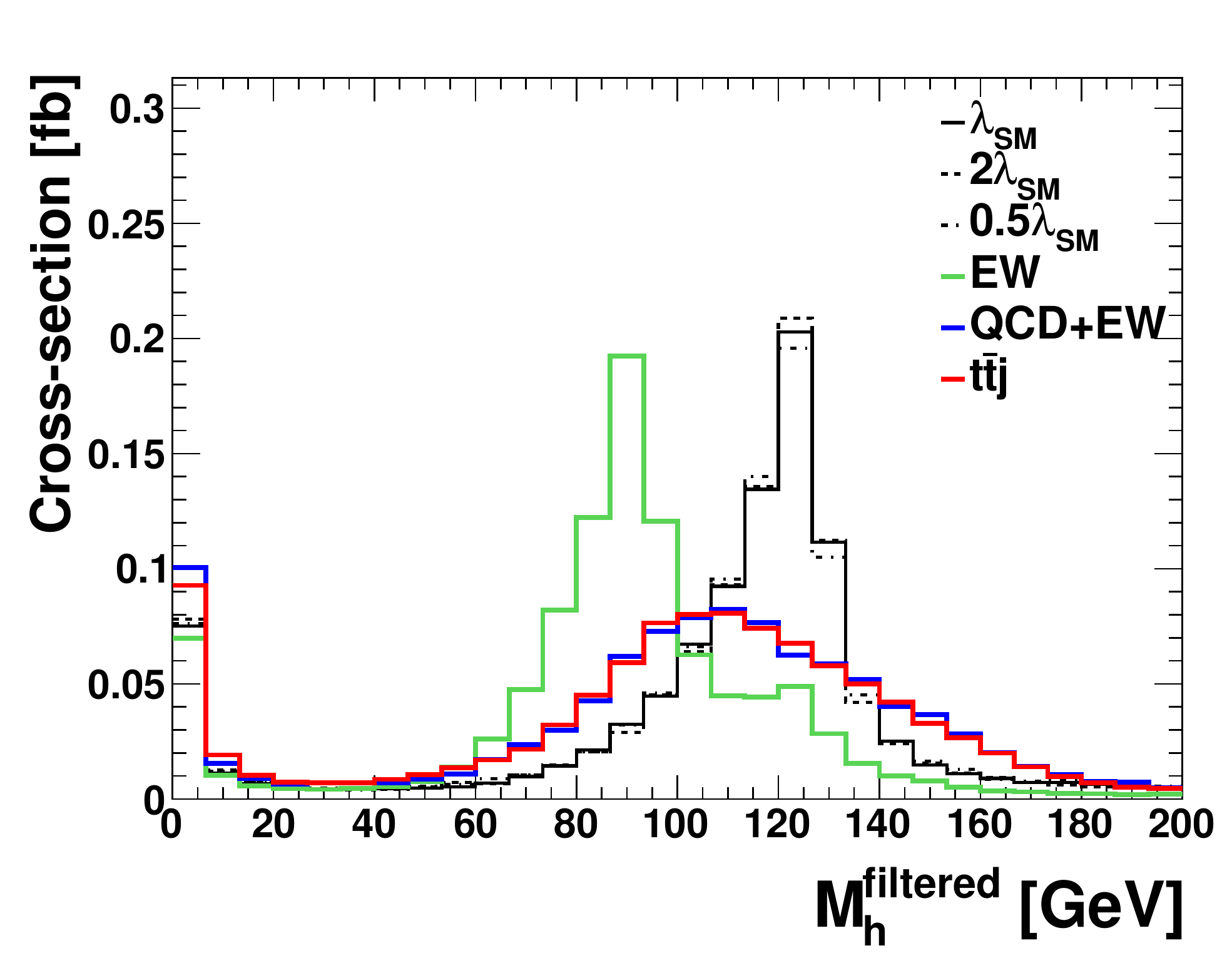}
\caption{Discriminating observables contributing to the boosted analysis of Sec.~\ref{sec:substruc}.}
\label{fig:taLtaH_substructure1}
\end{figure}
%%%%%%%%%%%%%%%%%%%%%%%%%%%%%%%

%%%%%%%%%%%%%%%%%%%%%%%%%%%%%%%
\begin{figure}[ht]
\centering
\includegraphics[scale=0.35]{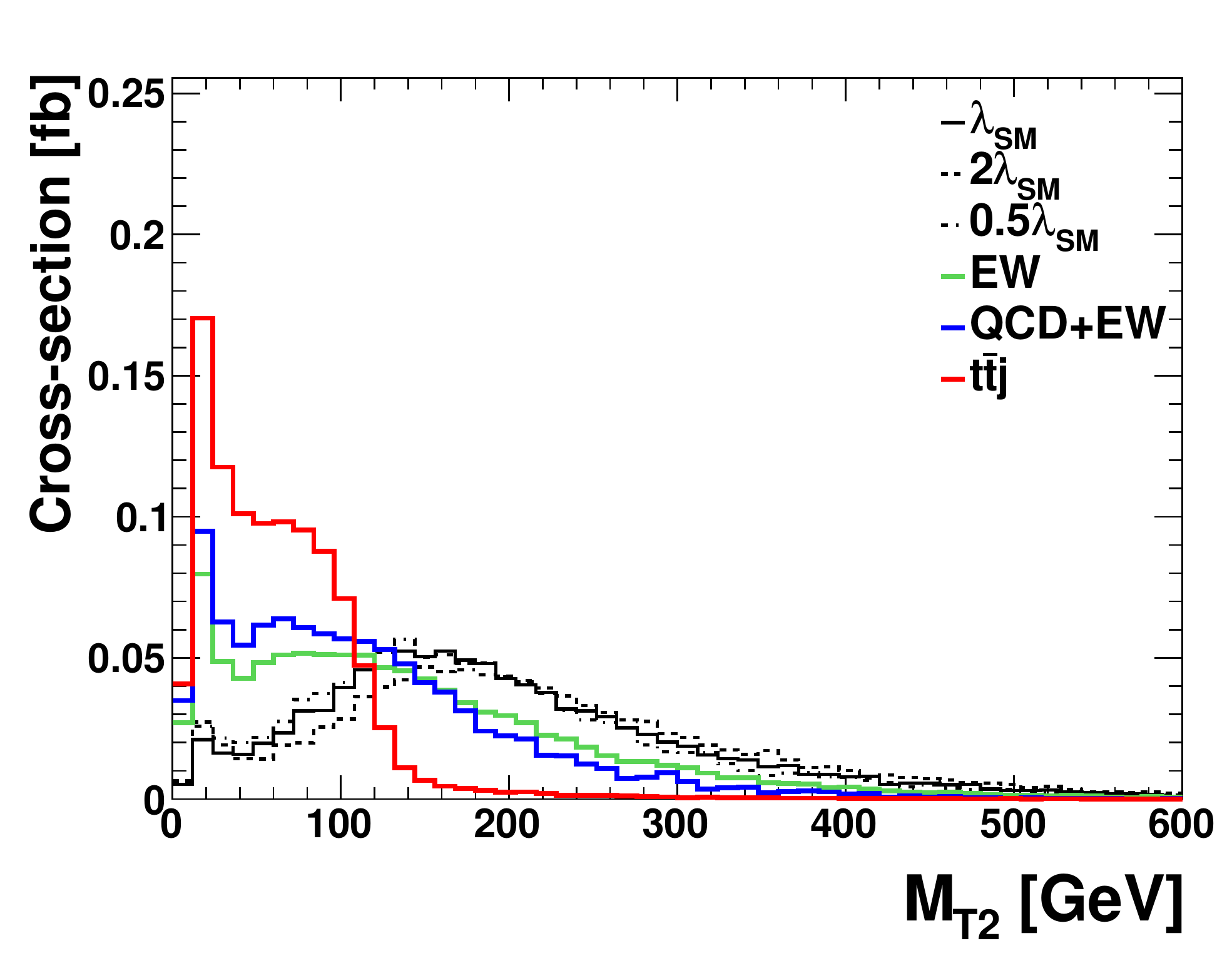}\hfill \includegraphics[scale=0.35]{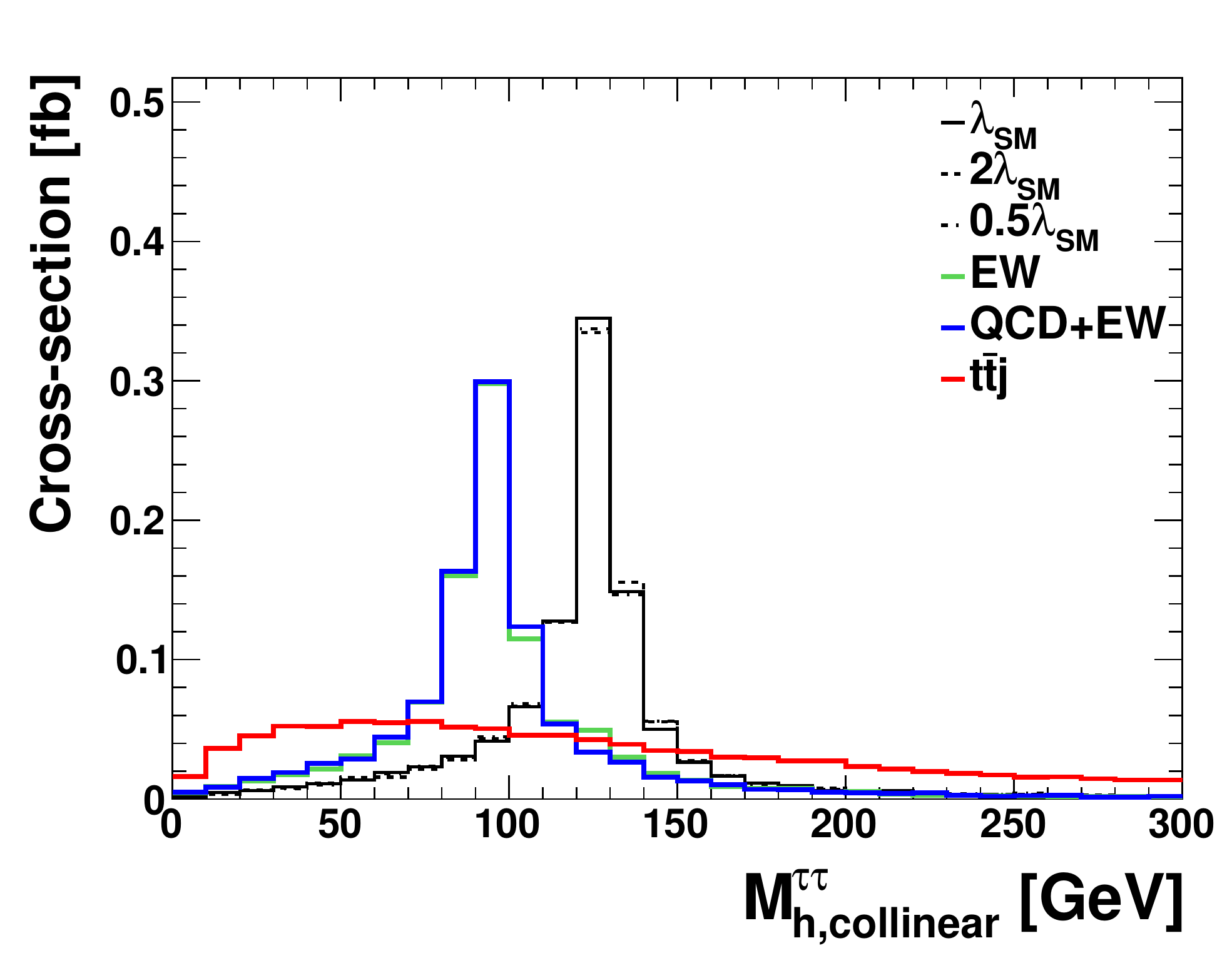}\\
\parbox{0.45\textwidth}{\includegraphics[scale=0.35]{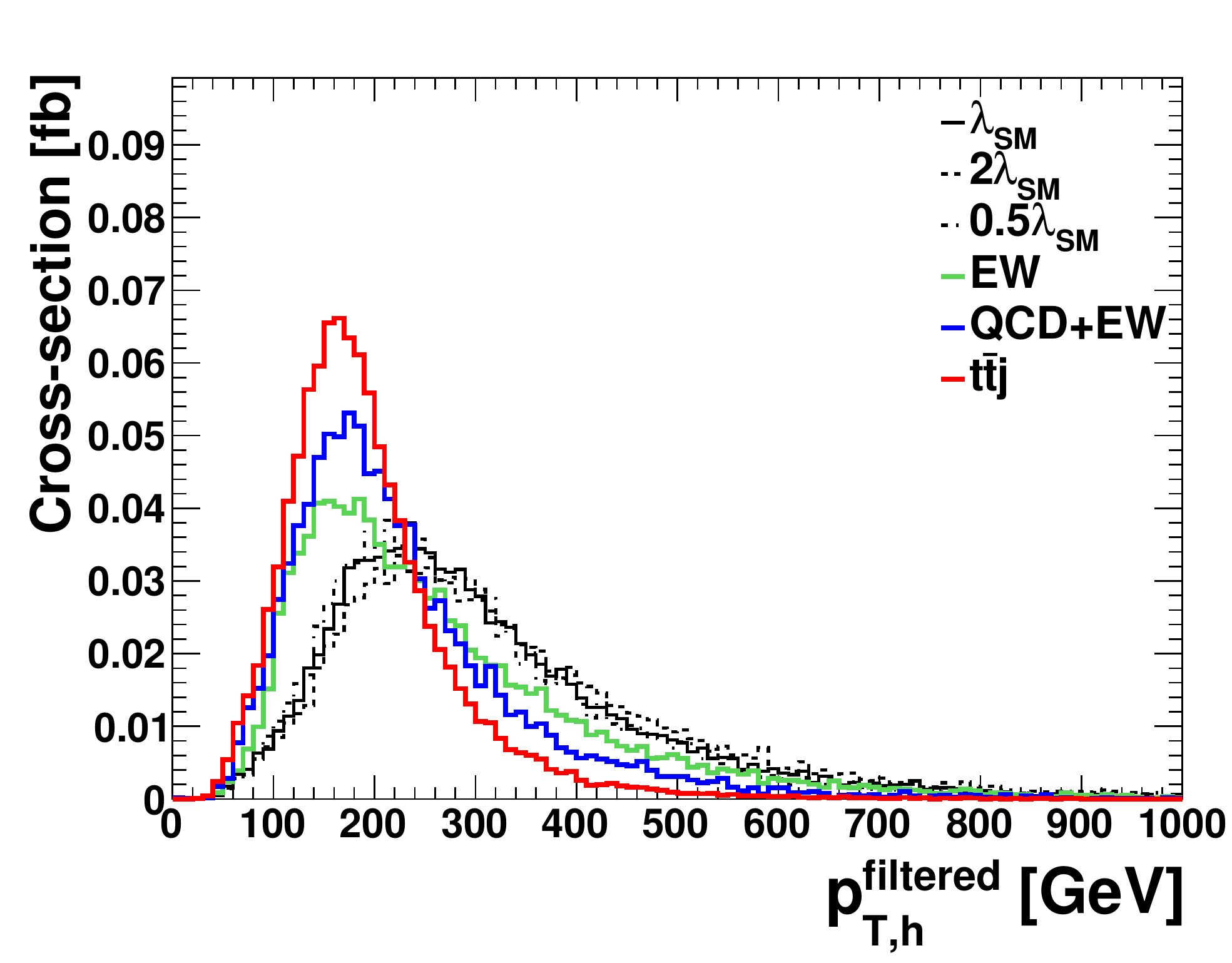}}
\hfill \parbox{0.45\textwidth}{\caption{\label{fig:taLtaH_substructure2} Discriminating observables contributing to the boosted analysis of Sec.~\ref{sec:substruc}.}}
\end{figure}
%%%%%%%%%%%%%%%%%%%%%%%%%%%%%%%

%%%%%%%%%%%%%%%%%%%%%%%%%%%%%%%
%\begin{figure}[ht]
%\centering
%\includegraphics[scale=0.65]{Figures/Fake_rate_vs_signal_acceptance.pdf}
%\caption{ROC curve of the analyses presented in this work.}
%\label{fig:taLtaH_substructure3}
%\end{figure}
%%%%%%%%%%%%%%%%%%%%%%%%%%%%%%%

To isolate this particular region, we change the analysis approach of Secs.~\ref{sec:2.2.1} and \ref{sec:2.2.2}. Before passing the events to the BDT we require at least two so-called fat jets of size $R=1.5$ and $p_T^{j} > 110~\text{GeV}$. One of these fat jets is required to contain displaced vertices associated with B mesons.
%\footnote{\checkthis{Through improvements of the calorimetry and trigger performance such information is likely to be available at a future hadron collider....}}.
We remove the jet constituents (that can contain leptons) and re-cluster the event along with our standard anti-kT choice. We then require either two isolated leptons ($\tau_\ell\tau_\ell$ cases) or one isolated lepton together with one $\tau$-tagged jet ($p_T>30~\text{GeV}$) using again a tagging efficiency of 70\% ($\tau_\ell\tau_h$ cases). All these objects are required to be in the central part of the detector $|\eta| <2.5$. Subsequently we apply substructure techniques to the jet containing displaced vertices following the by-now standard procedure of Ref.~\cite{Butterworth:2008iy} (we refer the reader for details to this publication and limit ourselves to quoting our choices of mass drop parameter 0.667 and $\sqrt{y}=0.3$). After jet-filtering we double-$b$ tag the two hardest subjets with an efficiency of 70\% (2\% mistag rate) and require the identified B-mesons to have $p_T>25~\text{GeV}$~\footnote{Furthermore, we also require light jets faking $b$-jets to have $p_T > 25$ GeV.}.
Finally, we require the leptons to be separated by $\Delta R(\ell \ell)>0.2$ in the $\tau_\ell\tau_\ell$ case. In the $\tau_h\tau_\ell$ case we require the lepton to be sufficiently well-separated from the hadronic tau $\Delta R(\ell,\tau_h)>0.4$.

We use the (jet-substructure) observables of Tab.~\ref{tab:subs} as BDT input~\footnote{Here also we obtain no change 
in sensitivity upon removing the redundant variables, \textit{viz.}, $p_{T,2}^{\textrm{filt}}, p_T^{\tau_{\textrm{vis},1}}, p_T^j, \Delta R (b_2 j), \Delta R (\tau_{\textrm{vis},1} j), \Delta R (b_1 j)$ and $\frac{p_T^{\tau_{\textrm{vis},2}}}{p_T^{\tau_{\textrm{vis},1}}}$.} (for a discussion of redundancies of the used observables see below). The signal vs. background discriminating power is shown in Figs.~\ref{fig:taLtaH_substructure1} and~\ref{fig:taLtaH_substructure2}. We can increase the sensitivity of the signal by using the collinear approximation outlined in Ref.~\cite{Elagin:2010aw} for the $\tau\tau$ pair.

%%%%%%%%%%%%%%%%%
\begin{table}[!t]
\centering
\begin{tabular}{| c  | l |}
\hline
observable & reconstructed object \\
\hline
\multirow{5}{*}{$p_T$} & 2 hardest filtered subjets\\
& 2 visible $\tau$ objects ($\tau_{\ell}$ or $\tau_h$) \\
& hardest non $b,\tau$-tagged jet \\
& reconstructed Higgs from filtered jets \\
& reconstructed Higgs from visible $\tau$ final states \\
\hline
\multirow{2}{*}{$p_T$ ratios} & 2 hardest filtered jets \\
& 2 visible $\tau$ final state objects \\
\hline
$m_{T2}$ & described in Eq.~\eqref{eq:mt2dec} \\
\hline
\multirow{5}{*}{$\Delta R$} & two hardest filtered subjets \\
& two visible $\tau$ objects ($\tau_{\ell} \tau_{\ell}$ or $\tau_{\ell} \tau_{h}$)  \\
& $b$-tagged jets and lepton or $\tau_h$ \\
& $b$-tagged jets and jet $j_1$ \\
& lepton or $\tau_h$ with jet $j_1$\\
\hline
$M_{\tau\tau}^{\text{col}}$ &  collinear approximation of $h\to \tau \tau$ mass \\
\hline
$M^{\text{filt}}$ & filtered $j_1$ and $j_2$ (and $j_3$ if present) \\
\hline
$M^{vis.}_{hh}$ & filtered jets and leptons (or lepton and $\tau_h$) \\
\hline
$\cancel{E}_T$ &  reduce sub-leading backgrounds \\
\hline
\multirow{2}{*}{$\Delta \phi$} & between visible $\tau$ final state objects and $\cancel{E}_T$\\
& between filtered jets system and $\ell \ell$ (or $\ell$ $\tau_h$) systems\\
\hline
$N_{\textrm{jets}}$ & number of anti-$k_T$ jets with $R=0.4$ \\
\hline
\end{tabular}
\caption{\label{tab:subs} Observables included to the boosted decision tree for the jet-substructure analysis of $pp\to hhj$, Sec.~\ref{sec:substruc}.}
\end{table}
%%%%%%%%%%%%%%%%%

The combined results are tabulated in Tab.~\ref{tab:subsres}. As can be seen, this approach retains larger signal and background cross sections compared to the fully-resolved approach that has a combined $S/B\simeq 0.08$. The sensitivity to $\kappa_\lambda$ is slightly more pronounced in the jet-substructure approach as expected. Together with the increased statistical control we can therefore constrain $\kappa_\lambda$ slightly more tightly (assuming again no systematic uncertainties)
\begin{alignat}{5}
\label{eq:boostedresult}
0.76 < \kappa_\lambda < 1.28 \quad & 3/\text{ab}\,, \\
0.92 < \kappa_\lambda < 1.08 \quad & 30/\text{ab}\,,
\end{alignat}
at 68\% confidence level using the identical CLs approach as above.

Before concluding this section we note that for our $b\bar b \tau^+ \tau^-$ analyses, the $S/B$ values are
10\% or more for the boosted combined ($\tau_{\ell} \tau_h + \tau_{\ell} \tau_{\ell}$) analysis and the resolved $\tau_l \tau_h$
analysis. For the $\tau_{\ell} \tau_{\ell}$ analysis however, we get $S/B$ below 5\%. Such values of $S/B$ are not
uncommon in Higgs analyses at the LHC. For example the $S/B$ in the inclusive $H\to\gamma\gamma$ search is 1/30, and
in the observation of $VH(\to b\bar{b})$, the $S/B$ is in the range of 1-2\%~\cite{Aaboud:2017xsd}, depending on the vector boson decay
mode. Ultimately, what counts is the precision with which the background rate can be determined. In our case, as in
the LHC examples given above, the background rate can be extracted directly from the data, using the sidebands of the
various kinematical distributions that we consider.

%%%%%%%%%%%%%%%%%%%%%%%%%%%%%%%
\begin{table}[!t]
\centering
\begin{footnotesize}
\begin{tabular}{| c  c | c | c | c | c || c | c |}
\hline
& signal  & QCD+EW  & EW  & $t\bar t j$  & tot. background & $S/B$ & $S/\sqrt{B}$, 30/ab \\
\hline
$\kappa_\lambda=0.5$ & 0.428 & \multirow{3}{*}{0.95} &  \multirow{3}{*}{0.27} &   \multirow{3}{*}{2.31}  &  \multirow{3}{*}{3.53} &  0.121 & 39.44 \\
$\kappa_\lambda=1$ & 0.363 & & & & & 0.103 & 33.44 \\
$\kappa_\lambda=2$ & 0.264 & & & & &  0.075 & 24.31 \\
\hline
\end{tabular}
\end{footnotesize}
\caption{\label{tab:subsres} Results of the boosted $pp\to hhj$ decay channels outlined in Sec.~\ref{sec:substruc} in femtobarns (numbers to the left of the double vertical lines) after an optimised cut on the BDT output. We include results for three different choices of the self-coupling within the $\kappa$ framework~\cite{Heinemeyer:2013tqa}, $\kappa_\lambda=\lambda/\lambda_{\text{SM}}$; BDT training is performed with $\lambda=\lambda_{\text{SM}}$.}
\end{table}
%%%%%%%%%%%%%%%%%%%%%%%%%%%%%%%

\subsection{Comments on cut-and-count experiments and redundancies}
\label{sec:comments}
A possible source of criticism of BDT based signal selection is that they cannot be straightforwardly mapped onto cut-and-count analyses, and the obtained signal region does not necessarily consist of connected physical phase space regions. In a busy collider environment with many competing processes and background rates that exceed the expected signal by orders of magnitude, multivariate methods are nevertheless very powerful tools that allow to extract information in various forms.\footnote{For instance, the recently reported evidence of $t\bar t h$ production~\cite{Aaboud:2017jvq} crucially relies on neural net analyses.}

The kinematics of $pp\to h h j$ is fully determined by five independent parameters. This raises the question whether the observed correlations of observables might allow us to consider subsets of the observables listed above. We investigate this by systematically removing correlated observables to trace their impact on our final sensitivity; we focus on the boosted selection as it shows the largest physics potential.

When removing observables which exhibit correlations of more that 70\%, we find our signal yields decreased in the percent range while the background (most notably $t\bar tj$) increases by $\gtrsim 15\%$. The impact on the signal, although small in size, is such that the $\kappa_\lambda$-dependence of the cross section becomes flatter. In total, focussing on observables with less than 70\% correlation therefore translates into constraints on the trilinear coupling $
0.89 < \kappa_\lambda < 1.28$ at 30/ab, which is clearly worse than the projection of Eq.~\eqref{eq:boostedresult}. Decreasing our correlation threshold to 60\%, we find our sensitivity even further decreased. This, together with a uniform relative importance of the observables for the BDT output score, indicates that the comprehensive list of observables indeed provides important discriminatory power, in particular when fighting against the large $t\bar t j$ background.

We can test the robustness of our analysis by comparing it against a more traditional cut-and-count approach. As part of the BDT analysis we can use the BDT's observable ranking to choose rectangular cuts in a particularly adapted way. From the cut-flow documented in Tab.~\ref{tab:cutncount}, we see, that we can reproduce the BDT $S/B$ sensitivity within a factor of two.

\begin{table}[!t]
\begin{footnotesize}
\begin{center}
\begin{tabular}{| c | c c c | c c c  |}
\hline
\multirow{2}{*}{cut} & \multicolumn{6}{c|}{cross section after cut [fb]} \\
&  $\kappa_\lambda=1$ & $\kappa_\lambda=0.5$ & $\kappa_\lambda=2$ & QCD+EW & EW & $t\bar t j$  \\
\hline
preselection &   0.86   &     1.09   &    0.56  &   11.73    & 2.20 &     4090.29 \\
$m_{T2} > 120$~GeV  &   0.65      &     0.78       &          0.45      &         4.65    &    1.10  &   300.68 \\
$\Delta\Phi(\tau_{\text{vis,2}}, \slashed{E}_T) < 1.5$ &   0.62   &        0.74           &      0.43        &       4.43   &     1.05   & 196.36 \\
$100~{\text{GeV}} < M_{\tau,\tau} < 150~{\text{GeV}}$ &  0.48     &      0.57      &           0.33     &          0.96   &     0.26  &  28.05 \\
$\Delta\Phi(\tau_{\text{vis,1}}, \slashed{E}_T) < 1.5$ &  0.47       &    0.56        &         0.32      &         0.92    &    0.25  &  21.75\\
$\Delta R(b_1 \tau_{\text{vis,1}}) > 0.8$ &  0.47     &      0.56         &        0.32       &        0.92   &     0.25  &  20.28 \\
$p_{T}(H_{\tau_{\text{vis,1}}\tau_{\text{vis,2}}}) > 60.0$~GeV  &  0.45   &        0.53      &           0.31     &          0.88   &     0.24  &  19.02 \\
$\Delta R(b_1 \tau_{\text{vis,2}}) > 0.8$ &      0.44     &      0.52            &     0.31    &           0.87     &   0.24  &  18.91�\\
$100~{\text{GeV}} < M_{\text{inv},\text{filt}} < 150~{\text{GeV}}$ &   0.32   &        0.38         &        0.22       &        0.43 &       0.06   & 5.78 \\
$\Delta R(b_1b_2) > 0.8$ &    0.32   &        0.38       &          0.22       &        0.43    &    0.06   & 5.46 \\
$\Delta R(\tau_{\text{vis,1}} \tau_{\text{vis,2}}) > 0.8$ &  0.31      &     0.37            &     0.22       &        0.42      &  0.06   & 5.15 \\
\hline
  & 0.36 &         0.44 &             0.26 &           0.95 &      0.27 &  2.31 \\
  \multicolumn{7}{|c|}{BDT performance [fb]} \\
\hline
\end{tabular}
\end{center}
\end{footnotesize}
\caption{\label{tab:cutncount} Comparison of an optimised cut-and-count analysis with our BDT analysis of Sec.~\ref{sec:substruc}. We optimise the selection to obtain a comparable signal yield after all analysis cuts. The order of the selection criteria reflects their relative impact on the BDT score.}
\end{table}
%%%%%%%%%%%%%%%%%%%%%%%%

%%%%%%%%%%%%%%%%%%%%%%%%
\section{The $jbbbb$ channel}
\label{sec:jbbbb}

Finally, we consider the $b \bar{b} b \bar{b} j$ channel for completeness. In order to compete with the large pure QCD background that contributes to this process and to trigger the event we need to consider very hard jets, $p_T^{j_1}\gtrsim 300~\text{GeV}$. For a more efficient background simulation, we therefore again generate the background events already with relatively hard cuts at the generator level. We choose the jet transverse momentum $p_T^j > 250$ GeV, the $\Delta R$ separation between bottom quarks and the light jet $\Delta R_{b,j}> 0.4$, bottom quark transverse momentum threshold $p_T^b > 15$ GeV, as well as bottom rapidity range $|\eta^b| < 3.0$. Furthermore, the jet rapidity range is restricted to $|\eta^j| < 5.0$ and we also require the bottom quarks to be separated in distance $\Delta R_{b,b} > 0.2$ as well as invariant mass $M_{b,b} > 30$ GeV.\footnote{Jets are defined as in Sec.~\ref{sec:2.2.1}.} For the signal, we only impose the generation level cut, $p_T^{j_1} > 200$ GeV. Throughout this part of the analysis, we will include a flat $b$-tagging efficiency of 70\% with mistag efficiency 2\%.

To account for QCD corrections we use again global $K$ factors as described above. In addition to the backgrounds discussed for the $\tau$ channels, we also need to include a pure QCD background leading to four final state $b$ quarks. The QCD corrections for this highly-involved final state are not available. We choose to use $K=1$. We note that this is consistent with the range of $K$ factors for inclusive 4 jet production discussed in Ref.~\cite{Mangano:2016jyj}.

\subsection{The resolved channel}
\label{sec:jbbbb_resolved}

The signal vs. background ratio is small for such inclusive selections. Therefore, in order to assess the sensitivity that can be reached in principle, we will again employ a multi-variate analysis strategy. Before passing the events to the multi-variate algorithm, we pre-select events according to the $hh+\text{jet}$ signal event topology.
For the resolved analysis we require 4 $b$-tagged jets and at least one hard non $b$-tagged jet with $p_T^{j} > 300$ GeV. The $b$-tagged jets are required to have a minimum $p_T$ of 30 GeV and need to fall inside the central detector region $|\eta^b| < 2.5$. All reconstructed objects need to be separated by $\Delta R >0.4$. Furthermore we define two masses: Firstly, $M_h^{\text{min},M}$ which is the reconstructed Higgs masses from pairing $b$-tagged jets close to the Higgs mass of 125~GeV. And secondly, $M_h^{\text{min},\Delta R}$ which follows from requiring that the first Higgs arises from the $b$-tagged jets with the smallest $\Delta R$ separation. We require that both $M_h^{\text{min},M},M_h^{\text{min},\Delta R}>30~\text{GeV}$. Finally we only use the Higgs bosons reconstructed upon utilising the minimum mass difference procedure.

Again we input a number of kinematic distributions to the BDT, detailed in Tab.~\ref{tab:bbbb}, the results are shown in Tab.~\ref{tab:bbbbres}. The signal vs. background ratio is extremely small, $O(10^{-3})$, leaving the analysis highly sensitive to systematic uncertainties with only little improvement possible using jet-substructure approaches.

%%%%%%%%%%%%%%%%%
\begin{table}[!t]
\centering
\begin{tabular}{| c  | l |}
\hline
observable & reconstructed object \\
\hline
\multirow{3}{*}{$p_T$} & 4 $b$-tagged jets \\
& hardest non $b$-tagged jet \\
& reconstructed $h\to b\bar b$ for both $M_h^{\text{min},M}$ definition \\
\hline
$p_T$ ratio & 4 $b$-tagged jets taken in pairs \\
\hline
\multirow{2}{*}{$\Delta R$} & $b$-tagged jets \\
& $b$-tagged jets and non-$b$-tagged jet \\
\hline
\multirow{1}{*}{$M$} & 4 $b$-tagged jets \\
\hline
$M_h^{\text{min},M}$ & see text\\
\hline
$\Delta \phi$  & between $h\to b\bar b$ for the $M_h^{\text{min},M}$ definition \\
\hline
\end{tabular}
\caption{\label{tab:bbbb} Observables included to the boosted decision tree for the fully-resolved 4 $b$-jet analysis of $pp\to hhj$, Sec.~\ref{sec:jbbbb}.}
\end{table}
%%%%%%%%%%%%%%%%%

%%%%%%%%%%%%%%%%%%%%%%%%%%%%%%%
\begin{table}[!t]
\centering
\begin{footnotesize}
\begin{tabular}{| c  c | c | c | c | c || c | c |}
\hline
& signal  & QCD  & QCD+EW  & EW & tot. background & $S/B\times 10^3$ & $S/\sqrt{B}$, 30/ab \\
\hline
$\kappa_\lambda=0.5$ & 0.252 & \multirow{3}{*}{41.67} &  \multirow{3}{*}{1.86} &   \multirow{3}{*}{0.13}  &  \multirow{3}{*}{43.66} &  5.8 & 6.61 \\
$\kappa_\lambda=1$ & 0.230 & & & & & 5.3 & 6.03 \\
$\kappa_\lambda=2$ & 0.160 & & & & &  3.6 &  4.18 \\
\hline
\end{tabular}
\end{footnotesize}
\caption{\label{tab:bbbbres} Results for the fully-resolved 4 $b$-jet analysis of $pp\to hhj$, Sec.~\ref{sec:jbbbb}. We include results for three different choices of the self-coupling within the $\kappa$ framework~\cite{Heinemeyer:2013tqa}, $\kappa_\lambda=\lambda/\lambda_{\text{SM}}$; BDT training is performed with $\lambda=\lambda_{\text{SM}}$. Numbers to the left of the double vertical lines are in femtobarns.}
\end{table}
%%%%%%%%%%%%%%%%%%%%%%%%%%%%%%%

\subsection{The boosted channel}
\label{sec:jbbbb_boosted}

We follow here the philosophy of Sec.~\ref{sec:substruc}, by exploiting the fact that most of the  sensitivity to the Higgs self-coupling comes from configurations where the di-Higgs system has a small invariant mass. This can be achieved by requiring the di-Higgs system to recoil against one or more high $p_T^{j}$ jets. If the Higgses have enough transverse momentum, their decay products, the $b \bar{b}$ pairs, will be collimated and eventually will be clustered as large radius jets. Such jets can be identified and disentangled from QCD jets with the use of standard substructure techniques.

Events are first pre-selected by requiring at least two central fat jets with parameter $R=0.8$ that contain at least two b-subjets. The fat jets are selected if $p_T^{j} > 300$ GeV and $|\eta^{j}| < 2.5$. We assume, as previously, a conservative $70\%$ $b$-tagging efficiency.  We further ask the di-fatjet pair to be sufficiently boosted, $p_T^{jj} > 250$ GeV, and the leading jet to have a $p_T^{j_1} > 400$ GeV. Finally, we require that $\Delta R(j_1,j_2) < 3.0$ as well as $(p_T^{j_1} - p_T^{j_2})/p_T^{jj} < 0.9$.

%%%%%%%%%%%%%%%%%%%%%%%%%%%%%%%%%%%%%%%%%%%%%%%%%%%%%%%%%%%%%
\begin{figure}[H]
\centering
\includegraphics[scale=0.35]{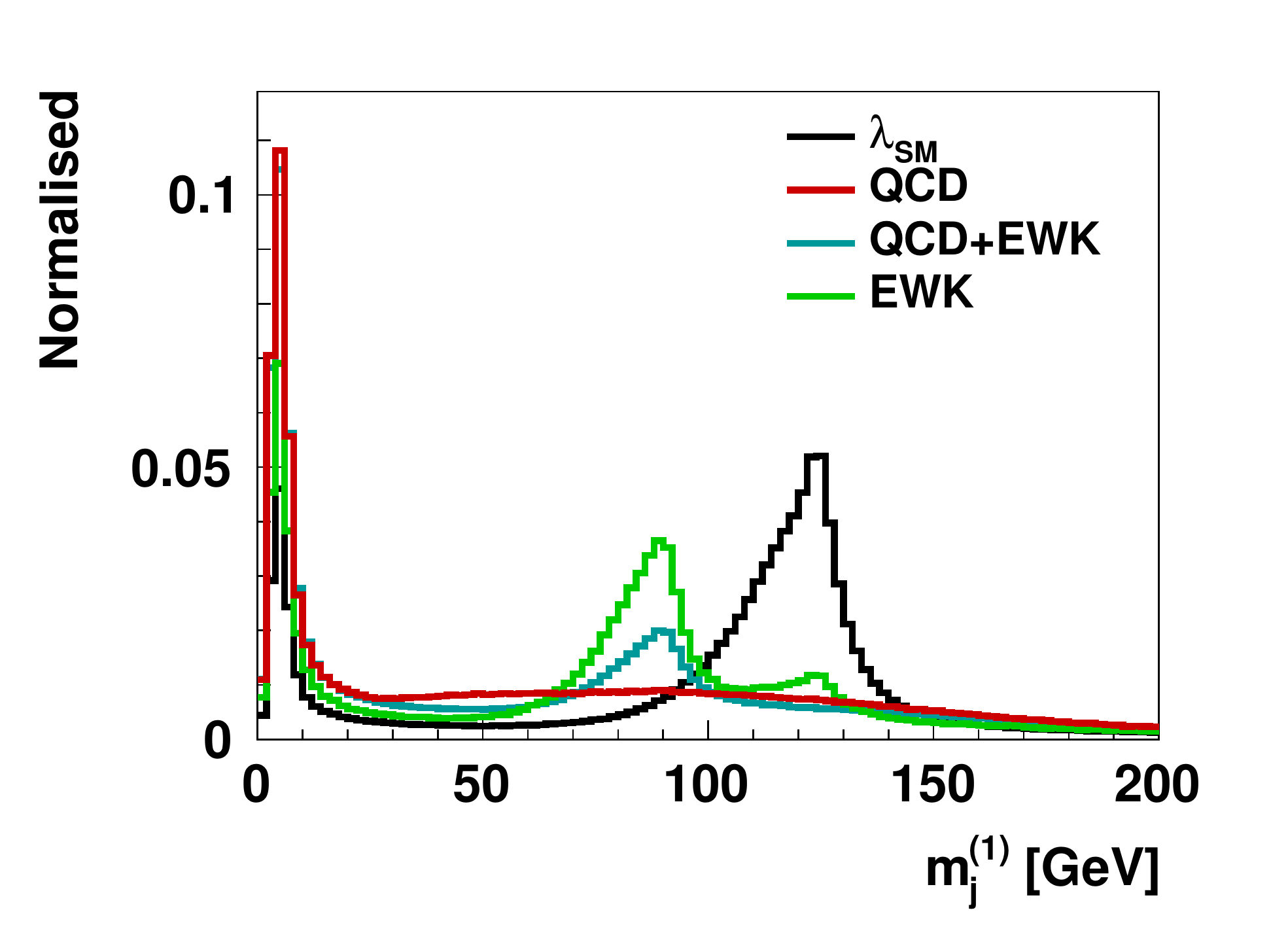}\hfill \includegraphics[scale=0.35]{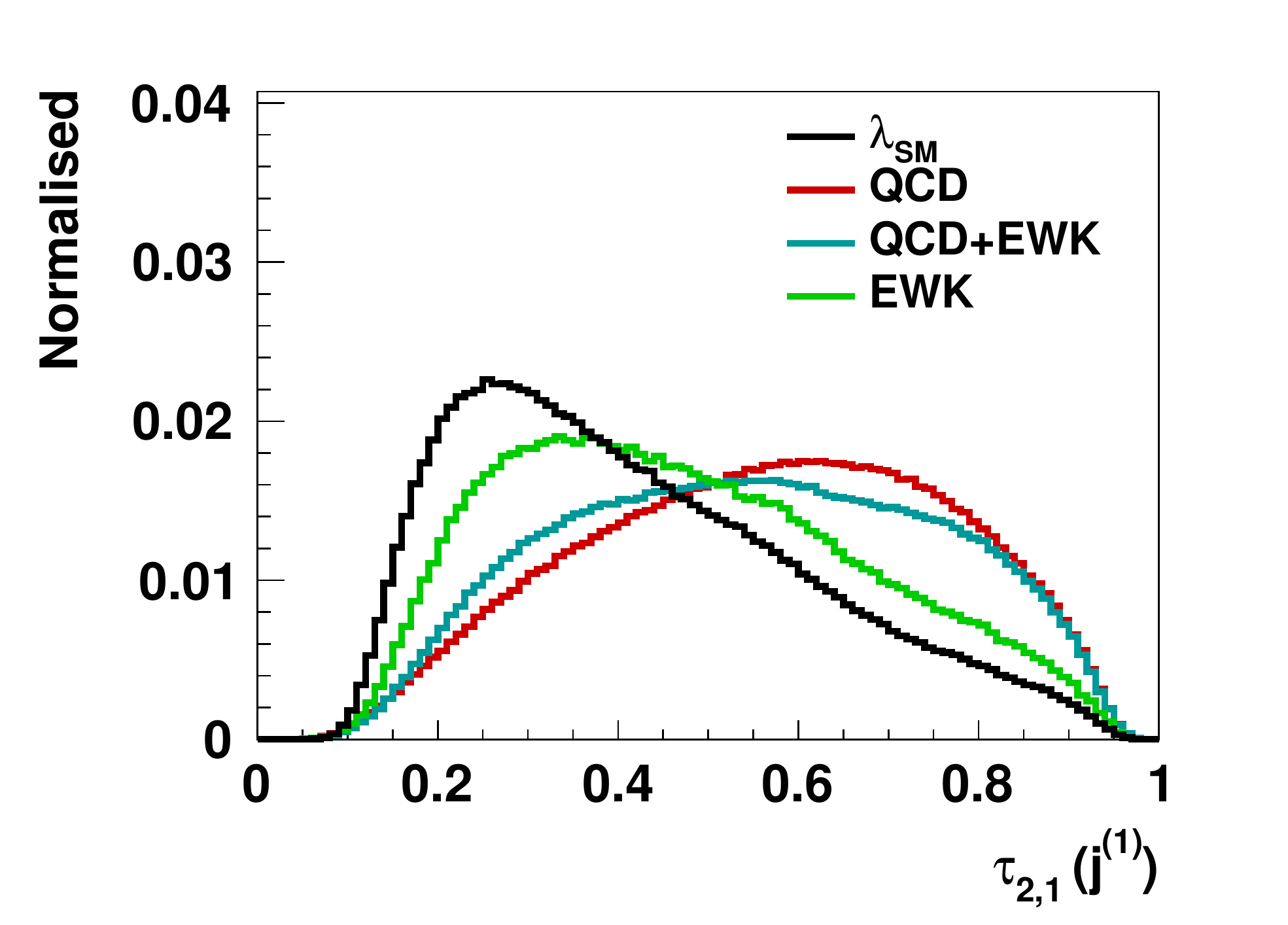}\\
{\caption{\label{fig:higgs_tag} Normalised differential soft-dropped mass $m_{SD}$ (left) and N-subjettiness ratio $\tau_{2,1}$ (right) for the leading reconstructed fat jet. The observables are used for the Higgs-jet tagging.}}
\end{figure}
%%%%%%%%%%%%%%%%%%%%%%%%%%%%%%%%%%%%%%%%%%%%%%%%%%%%%%%%%%%%%

The last steps of the event selection make use of jet-substructure observables and are designed to identify the collimated Higgs fat jets with high purity. The main background contribution is QCD $g \rightarrow b \bar{b}$ events, where configurations are dominated by soft and collinear splittings. The resulting jets are hence often characterized by one hard prong, as opposed to fat jets containing the Higgs decay products, that will feature a clear two-prong structure. The "2" versus "1" prong hypotheses of a jet can be tested with the $\tau_{2,1}$ observable~\cite{Thaler:2010tr}. Moreover Higgs jets typically have an invariant mass close to $m_H = 125$ GeV, as opposed to QCD jets that tend to have a small mass. QCD jets can therefore be rejected by requiring a soft-dropped mass $m_{SD}$~\cite{Larkoski:2014wba} of the order of the Higgs mass. These two observables are shown in Figure~\ref{fig:higgs_tag} for the leading reconstructed fat-jet.  The Higgs-jet tag consists in selecting jets with $\tau_{2,1} < 0.35$ and $100 < m_{SD} < 130$ GeV. This simple selection yields a tagging efficiency of 6\% and a mistag rate of 0.1\%. We apply the Higgs-jet tagging procedure to the two fat jets.

%%%%%%%%%%%%%%%%%%%%%%%%%%%%%%%
\begin{table}[!t]
\centering
\begin{footnotesize}
\begin{tabular}{| c  c | c | c | c | c || c | c |}
\hline
& signal & QCD & QCD+EW & EW & tot. background & $S/B\times 10^3$ & $S/\sqrt{B}$, 30/ab \\
\hline
$\kappa_\lambda=0.5$ & 0.094 & \multirow{3}{*}{4.3} & \multirow{3}{*}{0.1} & \multirow{3}{*}{0.003} & \multirow{3}{*}{4.4} & 20.8 & 7.67 \\
$\kappa_\lambda=1$ & 0.085 & & & & & 19.1 & 6.61 \\
$\kappa_\lambda=2$ & 0.071 & & & & & 16.2 & 5.85  \\
\hline
\end{tabular}
\end{footnotesize}
\caption{\label{tab:bbbb_boosted} Results for the boosted 4 $b$-jet analysis of $pp\to hhj$, Sec.~\ref{sec:jbbbb}. We include results for three different choices of the self-coupling within the $\kappa$ framework~\cite{Heinemeyer:2013tqa}, $\kappa_\lambda=\lambda/\lambda_{\text{SM}}$.  Numbers to the left of the double vertical lines are in femtobarns.}
\end{table}
%%%%%%%%%%%%%%%%%%%%%%%%%%%%%%%

The final results for the boosted analysis are summarized in Table~\ref{tab:bbbb_boosted}.
Although we find only a mild improvement on the significance compared to the resolved analysis, there is a clear improvement on the signal over background ratio $\sim 0.02$, allowing to better control background systematics.

%%%%%%%%%%%%%%%%%%%%%%%%%%%%%%%
\section{Summary and Conclusions}
\label{sec:summary}
%%%%%%%%%%%%%%%%%%%%%%%%%%%%%%%
Di-Higgs searches and their associated interpretation in terms of new, non-resonant physics are a key motivation for a future high-energy $pp$ collider. Recent analyses have mainly focused on direct $pp\to hh$ production, which has the shortcoming of back-to-back Higgs production generically accessing a phase space region with only limited sensitivity to the modifications of the trilinear Higgs coupling. This situation can be improved by accessing kinematical configurations where a collinear Higgs pair recoils against a hard jet, thus accessing small invariant masses $M_{hh}\simeq 2m_t$ over a broad range of final state kinematics. This is the region where modifications of the trilinear Higgs coupling are most pronounced.

In this work, we have focussed on this $hhj$ final state  at a 100 TeV collider. As exclusive final state cross sections are small, we focus in particular on the dominant $hh\to b\bar b b\bar b$ and $hh\to b\bar b \tau^+ \tau^-$ decay channels. Multi-Higgs final states suffer from small rates even in these dominant Higgs decay modes, which necessitates considering multivariate analysis techniques. We find that although the four $b$ final state is challenged by backgrounds with some opportunities to enhance sensitivity at large momenta, the $hh\to b\bar b \tau\tau$ final states provide a promising avenue to add significant sensitivity to the search for non-standard Higgs interactions. In particular, the hadronic tau decay channels which can be isolated with cutting-edge reconstruction techniques introduced by the CMS collaboration, drives the sensitivity. Relying on boosted final states, we show that $hhj$ production could in principle allow to constrain the Higgs self-coupling at the 8\% level at 30/ab (assuming no systematic uncertainties and other couplings to be SM-like).
This precision is thus worse than the $\sim 4\%$ result obtained for the inclusive $hh(\to b\bar{b}\gamma\gamma)$ channel shown in Ref.~\cite{Contino:2016spe}. Given the complexities of these analyses involving the Higgs self-coupling, we find it important that there be several independent modes to probe its value with a precision below the 10\% threshold. Furthermore, the different kinematical regimes probed by the $hh$ and the $hhj$ measurements could be sensitive in different ways to possible deviations from the SM expectations.
This motivates $pp\to hh j$ with semi-leptonic tau decays as an additional main search channel for modified Higgs physics.

%%%%%%%%%%%%%%%%%%%%%%%%%%%%%%%
\subsubsection*{Acknowledgements}
%%%%%%%%%%%%%%%%%%%%%%%%%%%%%%%
S.B. would like to thank Shilpi Jain for her immense help during various stages of this project. S.B. would also like to thank Biplob
Bhattacherjee and Amit Chakraborty for many helpful and insightful discussions. S.B. acknowledges the support of the Indo French LIA
THEP (Theoretical high Energy Physics) of the CNRS, France and is also supported by a Durham Junior Research Fellowship COFUNDed
between Durham University and the European Union under grant agreement number 609412. C.E. is supported by the IPPP Associateship scheme and by the UK Science and Technology Facilities Council (STFC) under grant ST/P000746/1.

\bibliography{draft.bbl}

\providecommand{\href}[2]{#2}\begingroup\raggedright\begin{thebibliography}{10}

\bibitem{CMS-PAS-SUS-16-052}
{\bf CMS Collaboration} Collaboration, {\it {Search for supersymmetry in events
  with at least one soft lepton, low jet multiplicity, and missing transverse
  momentum in proton-proton collisions at $\sqrt{s}=13~\mathrm{TeV}$}},  Tech.
  Rep. CMS-PAS-SUS-16-052, CERN, Geneva, 2017.

\bibitem{ATLAS-CONF-2017-034}
{\bf ATLAS Collaboration} Collaboration, {\it {Search for direct top squark
  pair production in final states with two leptons in $\sqrt{s} = 13$ TeV $pp$
  collisions with the ATLAS detector}},  Tech. Rep. ATLAS-CONF-2017-034, CERN,
  Geneva, May, 2017.

\bibitem{ATLAS-CONF-2016-101}
{\bf ATLAS Collaboration} Collaboration, {\it {Search for pair production of
  vector-like top partners in events with exactly one lepton and large missing
  transverse momentum in $\sqrt{s}=13$ TeV $pp$ collisions with the ATLAS
  detector}},  Tech. Rep. ATLAS-CONF-2016-101, CERN, Geneva, Sep, 2016.

\bibitem{CMS-PAS-B2G-16-005}
{\bf CMS Collaboration} Collaboration, {\it {Search for a vectorlike top
  partner produced through electroweak interaction and decaying to a top quark
  and a Higgs boson using boosted topologies in the all-hadronic final state}},
   Tech. Rep. CMS-PAS-B2G-16-005, CERN, Geneva, 2016.

\bibitem{Buchmuller:1985jz}
W.~Buchmuller and D.~Wyler, {\it {Effective Lagrangian Analysis of New
  Interactions and Flavor Conservation}},  {\em Nucl. Phys.} {\bf B268} (1986)
  621--653.

\bibitem{Hagiwara:1986vm}
K.~Hagiwara, R.~D. Peccei, D.~Zeppenfeld, and K.~Hikasa, {\it {Probing the Weak
  Boson Sector in $e^+ e^- \to W^+ W^-$}},  {\em Nucl. Phys.} {\bf B282} (1987)
  253.

\bibitem{Grzadkowski:2010es}
B.~Grzadkowski, M.~Iskrzynski, M.~Misiak, and J.~Rosiek, {\it {Dimension-Six
  Terms in the Standard Model Lagrangian}},  {\em JHEP} {\bf 10} (2010) 085,
  [\href{http://arxiv.org/abs/1008.4884}{{\tt arXiv:1008.4884}}].

\bibitem{Englert:2014uua}
C.~Englert, A.~Freitas, M.~M. Muhlleitner, T.~Plehn, M.~Rauch, M.~Spira, and
  K.~Walz, {\it {Precision Measurements of Higgs Couplings: Implications for
  New Physics Scales}},  {\em J. Phys.} {\bf G41} (2014) 113001,
  [\href{http://arxiv.org/abs/1403.7191}{{\tt arXiv:1403.7191}}].

\bibitem{Englert:2015hrx}
C.~Englert, R.~Kogler, H.~Schulz, and M.~Spannowsky, {\it {Higgs coupling
  measurements at the LHC}},  {\em Eur. Phys. J.} {\bf C76} (2016), no.~7 393,
  [\href{http://arxiv.org/abs/1511.05170}{{\tt arXiv:1511.05170}}].

\bibitem{Falkowski:2015jaa}
A.~Falkowski, M.~Gonzalez-Alonso, A.~Greljo, and D.~Marzocca, {\it {Global
  constraints on anomalous triple gauge couplings in effective field theory
  approach}},  {\em Phys. Rev. Lett.} {\bf 116} (2016), no.~1 011801,
  [\href{http://arxiv.org/abs/1508.00581}{{\tt arXiv:1508.00581}}].

\bibitem{Corbett:2015ksa}
T.~Corbett, O.~J.~P. Eboli, D.~Goncalves, J.~Gonzalez-Fraile, T.~Plehn, and
  M.~Rauch, {\it {The Higgs Legacy of the LHC Run I}},  {\em JHEP} {\bf 08}
  (2015) 156, [\href{http://arxiv.org/abs/1505.05516}{{\tt arXiv:1505.05516}}].

\bibitem{Corbett:2015mqf}
T.~Corbett, O.~J.~P. Eboli, D.~Goncalves, J.~Gonzalez-Fraile, T.~Plehn, and
  M.~Rauch, {\it {The Non-Linear Higgs Legacy of the LHC Run I}},
  \href{http://arxiv.org/abs/1511.08188}{{\tt arXiv:1511.08188}}.

\bibitem{Dicus:1987ic}
D.~A. Dicus, C.~Kao, and S.~S.~D. Willenbrock, {\it {Higgs Boson Pair
  Production From Gluon Fusion}},  {\em Phys. Lett.} {\bf B203} (1988)
  457--461.

\bibitem{Glover:1987nx}
E.~W.~N. Glover and J.~J. van~der Bij, {\it {Higgs boson pair production via
  gluon fusion}},  {\em Nucl. Phys.} {\bf B309} (1988) 282--294.

\bibitem{Djouadi:1999rca}
A.~Djouadi, W.~Kilian, M.~Muhlleitner, and P.~M. Zerwas, {\it {Production of
  neutral Higgs boson pairs at LHC}},  {\em Eur. Phys. J.} {\bf C10} (1999)
  45--49, [\href{http://arxiv.org/abs/hep-ph/9904287}{{\tt hep-ph/9904287}}].

\bibitem{Plehn:1996wb}
T.~Plehn, M.~Spira, and P.~M. Zerwas, {\it {Pair production of neutral Higgs
  particles in gluon-gluon collisions}},  {\em Nucl. Phys.} {\bf B479} (1996)
  46--64, [\href{http://arxiv.org/abs/hep-ph/9603205}{{\tt hep-ph/9603205}}].
  [Erratum: Nucl. Phys.B531,655(1998)].

\bibitem{Baur:2002rb}
U.~Baur, T.~Plehn, and D.~L. Rainwater, {\it {Measuring the Higgs boson self
  coupling at the LHC and finite top mass matrix elements}},  {\em Phys. Rev.
  Lett.} {\bf 89} (2002) 151801,
  [\href{http://arxiv.org/abs/hep-ph/0206024}{{\tt hep-ph/0206024}}].

\bibitem{Plehn:2005nk}
T.~Plehn and M.~Rauch, {\it {The quartic higgs coupling at hadron colliders}},
  {\em Phys. Rev.} {\bf D72} (2005) 053008,
  [\href{http://arxiv.org/abs/hep-ph/0507321}{{\tt hep-ph/0507321}}].

\bibitem{Kling:2016lay}
F.~Kling, T.~Plehn, and P.~Schichtel, {\it {Maximizing the significance in
  Higgs boson pair analyses}},  {\em Phys. Rev.} {\bf D95} (2017), no.~3
  035026, [\href{http://arxiv.org/abs/1607.07441}{{\tt arXiv:1607.07441}}].

\bibitem{Papaefstathiou:2015paa}
A.~Papaefstathiou and K.~Sakurai, {\it {Triple Higgs boson production at a 100
  TeV proton-proton collider}},  {\em JHEP} {\bf 02} (2016) 006,
  [\href{http://arxiv.org/abs/1508.06524}{{\tt arXiv:1508.06524}}].

\bibitem{Fuks:2017zkg}
B.~Fuks, J.~H. Kim, and S.~J. Lee, {\it {Scrutinizing the Higgs quartic
  coupling at a future 100 TeV proton–proton collider with taus and b-jets}},
   {\em Phys. Lett.} {\bf B771} (2017) 354--358,
  [\href{http://arxiv.org/abs/1704.04298}{{\tt arXiv:1704.04298}}].

\bibitem{Baur:2003gp}
U.~Baur, T.~Plehn, and D.~L. Rainwater, {\it {Probing the Higgs selfcoupling at
  hadron colliders using rare decays}},  {\em Phys. Rev.} {\bf D69} (2004)
  053004, [\href{http://arxiv.org/abs/hep-ph/0310056}{{\tt hep-ph/0310056}}].

\bibitem{Baur:2003gpa}
U.~Baur, T.~Plehn, and D.~L. Rainwater, {\it {Examining the Higgs boson
  potential at lepton and hadron colliders: A Comparative analysis}},  {\em
  Phys. Rev.} {\bf D68} (2003) 033001,
  [\href{http://arxiv.org/abs/hep-ph/0304015}{{\tt hep-ph/0304015}}].

\bibitem{Dolan:2012rv}
M.~J. Dolan, C.~Englert, and M.~Spannowsky, {\it {Higgs self-coupling
  measurements at the LHC}},  {\em JHEP} {\bf 10} (2012) 112,
  [\href{http://arxiv.org/abs/1206.5001}{{\tt arXiv:1206.5001}}].

\bibitem{ATL-PHYS-PUB-2017-001}
{\bf ATLAS Collaboration} Collaboration, {\it {Study of the double Higgs
  production channel $H(\rightarrow b\bar{b})H(\rightarrow \gamma\gamma)$ with
  the ATLAS experiment at the HL-LHC}},  Tech. Rep. ATL-PHYS-PUB-2017-001,
  CERN, Geneva, Jan, 2017.

\bibitem{CMS:2015nat}
{\bf CMS Collaboration} Collaboration, {\it {Higgs pair production at the High
  Luminosity LHC}},  Tech. Rep. CMS-PAS-FTR-15-002, CERN, Geneva, 2015.

\bibitem{Goertz:2013kp}
F.~Goertz, A.~Papaefstathiou, L.~L. Yang, and J.~Zurita, {\it {Higgs Boson
  self-coupling measurements using ratios of cross sections}},  {\em JHEP} {\bf
  06} (2013) 016, [\href{http://arxiv.org/abs/1301.3492}{{\tt
  arXiv:1301.3492}}].

\bibitem{Adhikary:2017jtu}
A.~Adhikary, S.~Banerjee, R.~K. Barman, B.~Bhattacherjee, and S.~Niyogi, {\it
  {Revisiting the non-resonant Higgs pair production at the HL-LHC}},
  \href{http://arxiv.org/abs/1712.05346}{{\tt arXiv:1712.05346}}.

\bibitem{Kim:2018uty}
J.~H. Kim, Y.~Sakaki, and M.~Son, {\it {Combined analysis of double Higgs
  production via gluon fusion at the HL-LHC in the effective field theory
  approach}},  \href{http://arxiv.org/abs/1801.06093}{{\tt arXiv:1801.06093}}.

\bibitem{Yao:2013ika}
W.~Yao, {\it {Studies of measuring Higgs self-coupling with $HH\rightarrow
  b\bar b \gamma\gamma$ at the future hadron colliders}},  in {\em
  {Proceedings, 2013 Community Summer Study on the Future of U.S. Particle
  Physics: Snowmass on the Mississippi (CSS2013): Minneapolis, MN, USA, July
  29-August 6, 2013}}, 2013.
\newblock \href{http://arxiv.org/abs/1308.6302}{{\tt arXiv:1308.6302}}.

\bibitem{Barr:2014sga}
A.~J. Barr, M.~J. Dolan, C.~Englert, D.~E. Ferreira~de Lima, and M.~Spannowsky,
  {\it {Higgs Self-Coupling Measurements at a 100 TeV Hadron Collider}},  {\em
  JHEP} {\bf 02} (2015) 016, [\href{http://arxiv.org/abs/1412.7154}{{\tt
  arXiv:1412.7154}}].

\bibitem{Azatov:2015oxa}
A.~Azatov, R.~Contino, G.~Panico, and M.~Son, {\it {Effective field theory
  analysis of double Higgs boson production via gluon fusion}},  {\em Phys.
  Rev.} {\bf D92} (2015), no.~3 035001,
  [\href{http://arxiv.org/abs/1502.00539}{{\tt arXiv:1502.00539}}].

\bibitem{He:2015spf}
H.-J. He, J.~Ren, and W.~Yao, {\it {Probing new physics of cubic Higgs boson
  interaction via Higgs pair production at hadron colliders}},  {\em Phys.
  Rev.} {\bf D93} (2016), no.~1 015003,
  [\href{http://arxiv.org/abs/1506.03302}{{\tt arXiv:1506.03302}}].

\bibitem{Contino:2016spe}
R.~Contino et~al., {\it {Physics at a 100 TeV pp collider: Higgs and EW
  symmetry breaking studies}},  {\em CERN Yellow Report} (2017), no.~3
  255--440, [\href{http://arxiv.org/abs/1606.09408}{{\tt arXiv:1606.09408}}].

\bibitem{deFlorian:2016spz}
{\bf LHC Higgs Cross Section Working Group} Collaboration, D.~de~Florian
  et~al., {\it {Handbook of LHC Higgs Cross Sections: 4. Deciphering the Nature
  of the Higgs Sector}},  \href{http://arxiv.org/abs/1610.07922}{{\tt
  arXiv:1610.07922}}.

\bibitem{twiki}
{\url{https://twiki.cern.ch/twiki/bin/view/LHCPhysics/LHCHXSWGHH}}.

\bibitem{Dawson:1998py}
S.~Dawson, S.~Dittmaier, and M.~Spira, {\it {Neutral Higgs boson pair
  production at hadron colliders: QCD corrections}},  {\em Phys. Rev.} {\bf
  D58} (1998) 115012, [\href{http://arxiv.org/abs/hep-ph/9805244}{{\tt
  hep-ph/9805244}}].

\bibitem{deFlorian:2013jea}
D.~de~Florian and J.~Mazzitelli, {\it {Higgs Boson Pair Production at
  Next-to-Next-to-Leading Order in QCD}},  {\em Phys. Rev. Lett.} {\bf 111}
  (2013) 201801, [\href{http://arxiv.org/abs/1309.6594}{{\tt
  arXiv:1309.6594}}].

\bibitem{Borowka:2016ehy}
S.~Borowka, N.~Greiner, G.~Heinrich, S.~Jones, M.~Kerner, J.~Schlenk,
  U.~Schubert, and T.~Zirke, {\it {Higgs Boson Pair Production in Gluon Fusion
  at Next-to-Leading Order with Full Top-Quark Mass Dependence}},  {\em Phys.
  Rev. Lett.} {\bf 117} (2016), no.~1 012001,
  [\href{http://arxiv.org/abs/1604.06447}{{\tt arXiv:1604.06447}}]. [Erratum:
  Phys. Rev. Lett.117,no.7,079901(2016)].

\bibitem{Shao:2013bz}
D.~Y. Shao, C.~S. Li, H.~T. Li, and J.~Wang, {\it {Threshold resummation
  effects in Higgs boson pair production at the LHC}},  {\em JHEP} {\bf 07}
  (2013) 169, [\href{http://arxiv.org/abs/1301.1245}{{\tt arXiv:1301.1245}}].

\bibitem{deFlorian:2015moa}
D.~de~Florian and J.~Mazzitelli, {\it {Higgs pair production at
  next-to-next-to-leading logarithmic accuracy at the LHC}},  {\em JHEP} {\bf
  09} (2015) 053, [\href{http://arxiv.org/abs/1505.07122}{{\tt
  arXiv:1505.07122}}].

\bibitem{Frederix:2014hta}
R.~Frederix, S.~Frixione, V.~Hirschi, F.~Maltoni, O.~Mattelaer, P.~Torrielli,
  E.~Vryonidou, and M.~Zaro, {\it {Higgs pair production at the LHC with NLO
  and parton-shower effects}},  {\em Phys. Lett.} {\bf B732} (2014) 142--149,
  [\href{http://arxiv.org/abs/1401.7340}{{\tt arXiv:1401.7340}}].

\bibitem{Maltoni:2014eza}
F.~Maltoni, E.~Vryonidou, and M.~Zaro, {\it {Top-quark mass effects in double
  and triple Higgs production in gluon-gluon fusion at NLO}},  {\em JHEP} {\bf
  11} (2014) 079, [\href{http://arxiv.org/abs/1408.6542}{{\tt
  arXiv:1408.6542}}].

\bibitem{deLima:2014dta}
D.~E. Ferreira~de Lima, A.~Papaefstathiou, and M.~Spannowsky, {\it {Standard
  model Higgs boson pair production in the ( $ b\overline{b} $ )( $
  b\overline{b} $ ) final state}},  {\em JHEP} {\bf 08} (2014) 030,
  [\href{http://arxiv.org/abs/1404.7139}{{\tt arXiv:1404.7139}}].

\bibitem{Behr:2015oqq}
J.~K. Behr, D.~Bortoletto, J.~A. Frost, N.~P. Hartland, C.~Issever, and
  J.~Rojo, {\it {Boosting Higgs pair production in the $b\bar{b}b\bar{b}$ final
  state with multivariate techniques}},  {\em Eur. Phys. J.} {\bf C76} (2016),
  no.~7 386, [\href{http://arxiv.org/abs/1512.08928}{{\tt arXiv:1512.08928}}].

\bibitem{Barr:2013tda}
A.~J. Barr, M.~J. Dolan, C.~Englert, and M.~Spannowsky, {\it {Di-Higgs final
  states augMT2ed -- selecting $hh$ events at the high luminosity LHC}},  {\em
  Phys. Lett.} {\bf B728} (2014) 308--313,
  [\href{http://arxiv.org/abs/1309.6318}{{\tt arXiv:1309.6318}}].

\bibitem{ATL-PHYS-PUB-2015-046}
{\it {Higgs Pair Production in the $H(\rightarrow \tau\tau)H(\rightarrow
  b\bar{b})$ channel at the High-Luminosity LHC}},  Tech. Rep.
  ATL-PHYS-PUB-2015-046, CERN, Geneva, Nov, 2015.

\bibitem{CMS-PAS-HIG-16-012}
{\bf CMS Collaboration} Collaboration, {\it {Search for non-resonant Higgs
  boson pair production in the $\mathrm{b\overline{b}}\tau^+\tau^-$ final
  state}},  Tech. Rep. CMS-PAS-HIG-16-012, CERN, Geneva, 2016.

\bibitem{Sirunyan:2017djm}
{\bf CMS} Collaboration, A.~M. Sirunyan et~al., {\it {Search for Higgs boson
  pair production in events with two bottom quarks and two tau leptons in
  proton–proton collisions at $\sqrt s$ =13TeV}},  {\em Phys. Lett.} {\bf
  B778} (2018) 101--127, [\href{http://arxiv.org/abs/1707.02909}{{\tt
  arXiv:1707.02909}}].

\bibitem{Alwall:2014hca}
J.~Alwall, R.~Frederix, S.~Frixione, V.~Hirschi, F.~Maltoni, O.~Mattelaer,
  H.~S. Shao, T.~Stelzer, P.~Torrielli, and M.~Zaro, {\it {The automated
  computation of tree-level and next-to-leading order differential cross
  sections, and their matching to parton shower simulations}},  {\em JHEP} {\bf
  07} (2014) 079, [\href{http://arxiv.org/abs/1405.0301}{{\tt
  arXiv:1405.0301}}].

\bibitem{Frixione:2007zp}
S.~Frixione, E.~Laenen, P.~Motylinski, and B.~R. Webber, {\it {Angular
  correlations of lepton pairs from vector boson and top quark decays in Monte
  Carlo simulations}},  {\em JHEP} {\bf 04} (2007) 081,
  [\href{http://arxiv.org/abs/hep-ph/0702198}{{\tt hep-ph/0702198}}].

\bibitem{Artoisenet:2012st}
P.~Artoisenet, R.~Frederix, O.~Mattelaer, and R.~Rietkerk, {\it {Automatic
  spin-entangled decays of heavy resonances in Monte Carlo simulations}},  {\em
  JHEP} {\bf 03} (2013) 015, [\href{http://arxiv.org/abs/1212.3460}{{\tt
  arXiv:1212.3460}}].

\bibitem{Sjostrand:2014zea}
T.~Sj{\"{o}}strand, S.~Ask, J.~R. Christiansen, R.~Corke, N.~Desai, P.~Ilten,
  S.~Mrenna, S.~Prestel, C.~O. Rasmussen, and P.~Z. Skands, {\it {An
  Introduction to PYTHIA 8.2}},  {\em Comput. Phys. Commun.} {\bf 191} (2015)
  159--177, [\href{http://arxiv.org/abs/1410.3012}{{\tt arXiv:1410.3012}}].

\bibitem{Binoth:2009wk}
T.~Binoth, T.~Gleisberg, S.~Karg, N.~Kauer, and G.~Sanguinetti, {\it {NLO QCD
  corrections to ZZ+ jet production at hadron colliders}},  {\em Phys. Lett.}
  {\bf B683} (2010) 154--159, [\href{http://arxiv.org/abs/0911.3181}{{\tt
  arXiv:0911.3181}}].

\bibitem{Campbell:2002tg}
J.~M. Campbell and R.~K. Ellis, {\it {Next-to-leading order corrections to
  $W^+$ 2 jet and $Z^+$ 2 jet production at hadron colliders}},  {\em Phys.
  Rev.} {\bf D65} (2002) 113007,
  [\href{http://arxiv.org/abs/hep-ph/0202176}{{\tt hep-ph/0202176}}].

\bibitem{Bevilacqua:2015qha}
G.~Bevilacqua, H.~B. Hartanto, M.~Kraus, and M.~Worek, {\it {Top Quark Pair
  Production in Association with a Jet with Next-to-Leading-Order QCD Off-Shell
  Effects at the Large Hadron Collider}},  {\em Phys. Rev. Lett.} {\bf 116}
  (2016), no.~5 052003, [\href{http://arxiv.org/abs/1509.09242}{{\tt
  arXiv:1509.09242}}].

\bibitem{Lester:1999tx}
C.~G. Lester and D.~J. Summers, {\it {Measuring masses of semiinvisibly
  decaying particles pair produced at hadron colliders}},  {\em Phys. Lett.}
  {\bf B463} (1999) 99--103, [\href{http://arxiv.org/abs/hep-ph/9906349}{{\tt
  hep-ph/9906349}}].

\bibitem{Baringer:2011nh}
P.~Baringer, K.~Kong, M.~McCaskey, and D.~Noonan, {\it {Revisiting
  Combinatorial Ambiguities at Hadron Colliders with $M_{T2}$}},  {\em JHEP}
  {\bf 10} (2011) 101, [\href{http://arxiv.org/abs/1109.1563}{{\tt
  arXiv:1109.1563}}].

\bibitem{Cacciari:2008gp}
M.~Cacciari, G.~P. Salam, and G.~Soyez, {\it {The Anti-k(t) jet clustering
  algorithm}},  {\em JHEP} {\bf 04} (2008) 063,
  [\href{http://arxiv.org/abs/0802.1189}{{\tt arXiv:0802.1189}}].

\bibitem{Cacciari:2011ma}
M.~Cacciari, G.~P. Salam, and G.~Soyez, {\it {FastJet User Manual}},  {\em Eur.
  Phys. J.} {\bf C72} (2012) 1896, [\href{http://arxiv.org/abs/1111.6097}{{\tt
  arXiv:1111.6097}}].

\bibitem{ATLAS:2012ima}
{\bf ATLAS} Collaboration, {\it {Measurement of the b-tag Efficiency in a
  Sample of Jets Containing Muons with 5 fb−1 of Data from the ATLAS
  Detector}}, .

\bibitem{Aaboud:2017rss}
{\bf ATLAS} Collaboration, M.~Aaboud et~al., {\it {Search for the Standard
  Model Higgs boson produced in association with top quarks and decaying into a
  $b\bar{b}$ pair in $pp$ collisions at $\sqrt{s}$ = 13 TeV with the ATLAS
  detector}},  \href{http://arxiv.org/abs/1712.08895}{{\tt arXiv:1712.08895}}.

\bibitem{Hocker:2007ht}
A.~Hocker et~al., {\it {TMVA - Toolkit for Multivariate Data Analysis}},  {\em
  PoS} {\bf ACAT} (2007) 040, [\href{http://arxiv.org/abs/physics/0703039}{{\tt
  physics/0703039}}].

\bibitem{Heinemeyer:2013tqa}
{\bf LHC Higgs Cross Section Working Group} Collaboration, J.~R. Andersen
  et~al., {\it {Handbook of LHC Higgs Cross Sections: 3. Higgs Properties}},
  \href{http://arxiv.org/abs/1307.1347}{{\tt arXiv:1307.1347}}.

\bibitem{Cadamuro:2015lbd}
{\bf CMS} Collaboration, L.~Cadamuro, {\it {The CMS Level-1 Tau algorithm for
  the LHC Run II}},  {\em PoS} {\bf EPS-HEP2015} (2015) 226.

\bibitem{CMS-DP-2015-009}
{\bf CMS Collaboration} Collaboration, {\it {L1 calorimeter trigger upgrade:
  tau performance}}, .

\bibitem{Mastrolorenzo:2016dyo}
L.~Mastrolorenzo, {\it {The CMS Level-1 Tau identification algorithm for the
  LHC Run II}},  {\em Nucl. Part. Phys. Proc.} {\bf 273-275} (2016) 2518--2520.

\bibitem{Junk:1999kv}
T.~Junk, {\it {Confidence level computation for combining searches with small
  statistics}},  {\em Nucl. Instrum. Meth.} {\bf A434} (1999) 435--443,
  [\href{http://arxiv.org/abs/hep-ex/9902006}{{\tt hep-ex/9902006}}].

\bibitem{Read:2000ru}
A.~L. Read, {\it {Modified frequentist analysis of search results (The CL(s)
  method)}},  in {\em {Workshop on confidence limits, CERN, Geneva,
  Switzerland, 17-18 Jan 2000: Proceedings}}, pp.~81--101, 2000.

\bibitem{Read:2002hq}
A.~L. Read, {\it {Presentation of search results: The CL(s) technique}},  {\em
  J. Phys.} {\bf G28} (2002) 2693--2704. [,11(2002)].

\bibitem{Butterworth:2008iy}
J.~M. Butterworth, A.~R. Davison, M.~Rubin, and G.~P. Salam, {\it {Jet
  substructure as a new Higgs search channel at the LHC}},  {\em Phys. Rev.
  Lett.} {\bf 100} (2008) 242001, [\href{http://arxiv.org/abs/0802.2470}{{\tt
  arXiv:0802.2470}}].

\bibitem{Elagin:2010aw}
A.~Elagin, P.~Murat, A.~Pranko, and A.~Safonov, {\it {A New Mass Reconstruction
  Technique for Resonances Decaying to di-tau}},  {\em Nucl. Instrum. Meth.}
  {\bf A654} (2011) 481--489, [\href{http://arxiv.org/abs/1012.4686}{{\tt
  arXiv:1012.4686}}].

\bibitem{Aaboud:2017xsd}
{\bf ATLAS} Collaboration, M.~Aaboud et~al., {\it {Evidence for the $ H\to
  b\overline{b} $ decay with the ATLAS detector}},  {\em JHEP} {\bf 12} (2017)
  024, [\href{http://arxiv.org/abs/1708.03299}{{\tt arXiv:1708.03299}}].

\bibitem{Aaboud:2017jvq}
{\bf ATLAS} Collaboration, M.~Aaboud et~al., {\it {Evidence for the associated
  production of the Higgs boson and a top quark pair with the ATLAS detector}},
   {\em Submitted to: Phys. Rev. D} (2017)
  [\href{http://arxiv.org/abs/1712.08891}{{\tt arXiv:1712.08891}}].

\bibitem{Mangano:2016jyj}
M.~L. Mangano et~al., {\it {Physics at a 100 TeV pp Collider: Standard Model
  Processes}},  {\em CERN Yellow Report} (2017), no.~3 1--254,
  [\href{http://arxiv.org/abs/1607.01831}{{\tt arXiv:1607.01831}}].

\bibitem{Thaler:2010tr}
J.~Thaler and K.~Van~Tilburg, {\it {Identifying Boosted Objects with
  N-subjettiness}},  {\em JHEP} {\bf 03} (2011) 015,
  [\href{http://arxiv.org/abs/1011.2268}{{\tt arXiv:1011.2268}}].

\bibitem{Larkoski:2014wba}
A.~J. Larkoski, S.~Marzani, G.~Soyez, and J.~Thaler, {\it {Soft Drop}},  {\em
  JHEP} {\bf 05} (2014) 146, [\href{http://arxiv.org/abs/1402.2657}{{\tt
  arXiv:1402.2657}}].

\end{thebibliography}\endgroup
\end{document}